\documentclass{article}

\usepackage{arxiv}

\usepackage[utf8]{inputenc} 
\usepackage[T1]{fontenc}    
\usepackage{hyperref}       
\usepackage{url}            
\usepackage{booktabs}       
\usepackage{amsfonts}       
\usepackage{nicefrac}       
\usepackage{microtype}      
\usepackage{lipsum}		
\usepackage{graphicx}
\usepackage{natbib}
\usepackage{doi}
\usepackage{caption}
\usepackage{subcaption}
\usepackage{algorithm}
\usepackage{algorithmic}

\usepackage{url}
\usepackage{xcolor}
\definecolor{newcolor}{rgb}{.8,.349,.1}

\definecolor{myR}{rgb}{0.75, 0.0, 0.0} 
\definecolor{myG}{rgb}{0.0, 0.75, 0.0} 
\definecolor{myB}{rgb}{0.0, 0.0, 0.75}

\title{Medial Parametrization of Arbitrary \\Planar Compact  Domains with Dipoles}%

\author{ \href{https://orcid.org/0000-0001-6434-2577}{\includegraphics[scale=0.06]{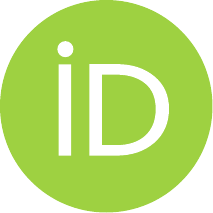}\hspace{1mm}Vinayak Krishnamurthy} \thanks{Joint with Computer Science and Engineering.}\\
	Mechanical Engineering, \\ Texas A\&M University, College Station, TX, 77831\\
	\texttt{vinayak@tamu.edu} \\
     \And
     \href{https://orcid.org/0000-0003-3618-4166}{\includegraphics[scale=0.06]{orcid.pdf}\hspace{1mm}Ergun Akleman}\thanks{Joint with Computer Science and Engineering.} \\
	Visual Computing \& Computational Media,\\ Texas A\&M University, College Station, TX, 77831\\
	\texttt{ergun@tamu.edu} \\
}

\renewcommand{\shorttitle}{Medial Parameterization with Dipoles}

\hypersetup{
pdftitle={A template for the arxiv style},
pdfsubject={q-bio.NC, q-bio.QM},
pdfauthor={David S.~Hippocampus, Elias D.~Striatum},
pdfkeywords={First keyword, Second keyword, More},
}

\begin{document}
\maketitle

\begin{abstract}
We present medial parametrization, a new approach to parameterize any compact planar domain bounded by a set of simple closed curves. The basic premise behind our proposed approach is to use two close Voronoi sites, which we call dipoles, to simultaneously construct and reconstruct an approximate piecewise-linear version of the original boundary and medial axis through Voronoi tessellation. 
The boundaries and medial axes of such planar compact domains, \textit{ in conjunction with}, offer a natural way to describe the interior of the domain. In fact, any compact planar domain homeomorphic to a compact unit circular disk admits a parameterization isomorphic to the $(r,\theta)$ or polar parametrization of the disk. Specifically, the medial axis and the boundary generalize the radial ($r$) and angular ($\theta$) parameters, respectively. In this paper, we present a simple algorithm that puts these principles into practice. The algorithm is based on the simultaneous re-creation of the boundaries of the domain and its medial axis using Voronoi tessellation. This simultaneous re-creation provides partitions of the domain into a set of ``skinny'' convex polygons wherein each polygon is essentially a subset of the medial edges (which we call the spine) connected to the boundary through exactly two straight edges (which we call limbs). It is this unique structure that enables us to convert the original Voronoi tessellation into quadrilaterals and triangles (at the poles of the medial axis) \textit{neatly ordered along the boundaries} of the domain, thereby allowing proper parametrization of the domain. Our approach is agnostic to the number of holes and disconnected components bounding the domain. We investigate the efficacy of our concept and algorithm through several examples. 
\end{abstract}


\begin{figure*}
  \centering
    \begin{subfigure}[t]{0.45\linewidth}
\includegraphics[width=\linewidth]{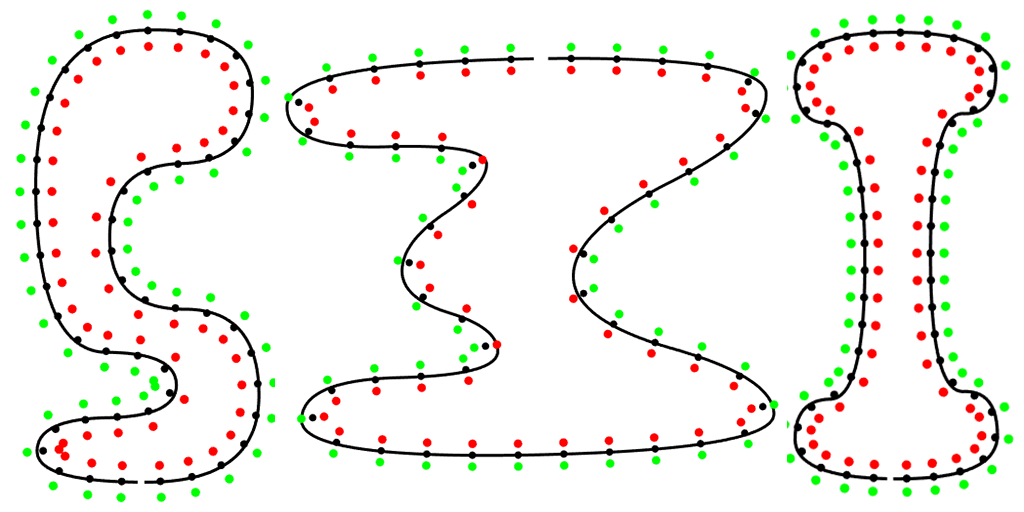}
\caption{Voronoi Sites: Dipoles.  }
\label{fig:Results/SMI/00}
    \end{subfigure}
        \hfill
    \begin{subfigure}[t]{0.45\linewidth}
\includegraphics[width=\linewidth]{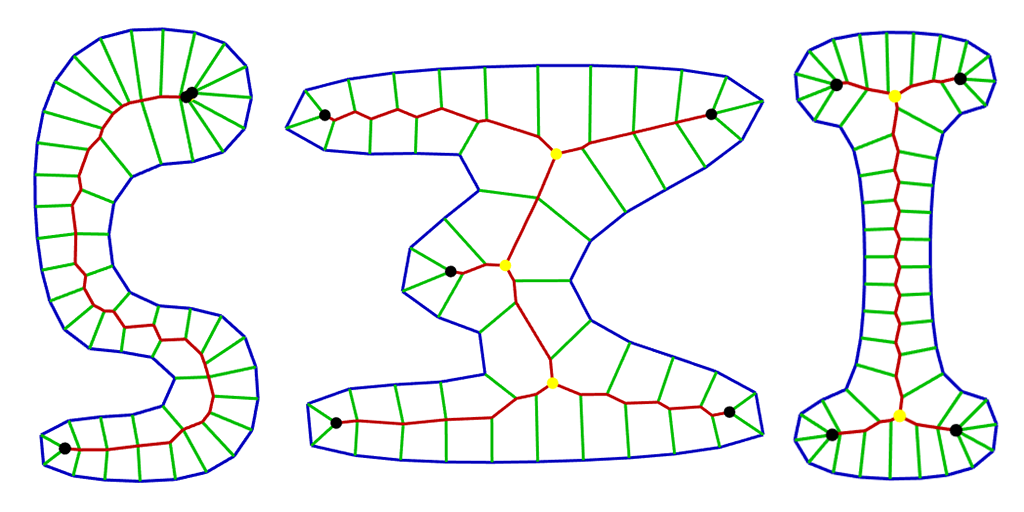}
\caption{Voronoi decomposition.  }
\label{fig:Results/SMI/01}
    \end{subfigure}
        \hfill
    \begin{subfigure}[t]{0.45\linewidth}
\includegraphics[width=\linewidth]{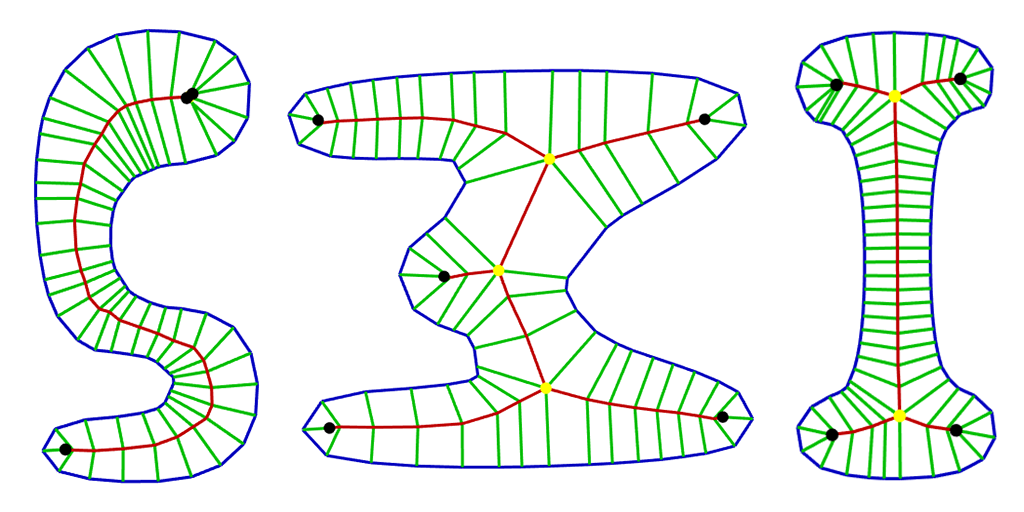}
\caption{Remeshing.  }
\label{fig:Results/SMI/02}
    \end{subfigure}
        \hfill
    \begin{subfigure}[t]{0.45\linewidth}
\includegraphics[width=\linewidth]{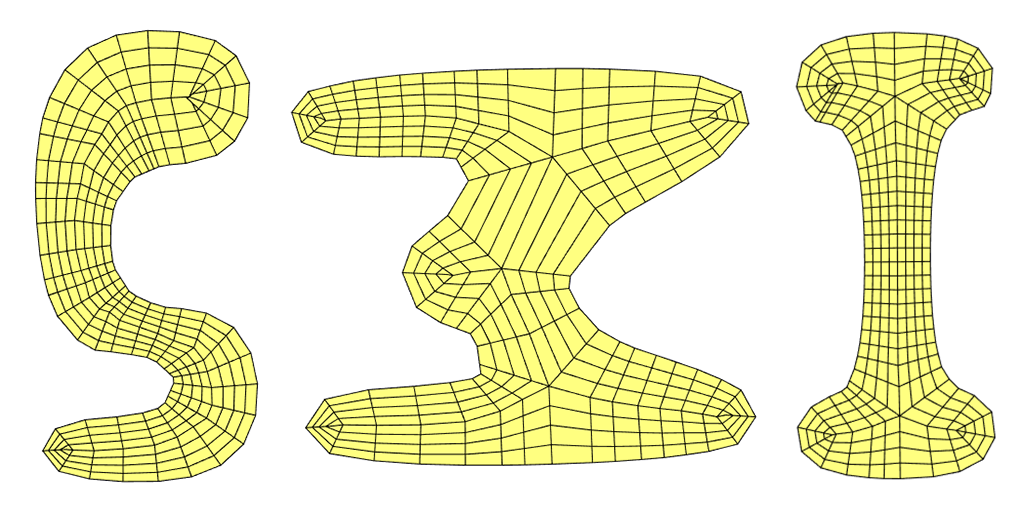}
\caption{Parametrization.  }
\label{fig:Results/SMI/03}
    \end{subfigure}
        \hfill
    \caption{An example that demonstrates the whole process for a domain that consists of three disconnected regions. }
    \label{fig:SMI}
\end{figure*}

\section{Introduction}

\subsection{Context \& Motivation}
Parametrization is fundamental to several graphics, geometric modeling, meshing, vision, and other geometric computing applications. As a result, it is also an extensively studied problem in literature. Most work on parametrization specifically deals with the problem of ``\textit{mesh parametrization}'' where the idea is to map a piece-wise linear representation of a given surface to some standardized space such as a unit square or a sphere~\cite{Sheffer2006}. Therefore, these problems are rather inverse parametrization problems (given a point in a domain, determine the parameters) and are useful for practical applications. 

When dealing with the forward problem however (given a domain, determine ways to sample points in it), we generally find ourselves resorting to domains that can be directly described by parametric equations for their constitutive curves. It is therefore not surprising that early efforts in computer-aided geometric design were mostly tensor product surfaces~\cite{FARIN20021}. The reason is simple: surfaces generated by sweeping a curve along another curve carry forward the curve's parametrization, and therefore the order of points, onto the surface. In contrast, the problem of finding a way to parameterize an arbitrarily shaped domain defined only by its boundaries is a hard problem both geometrically and topologically. 

\subsection{Basis \& Rationale}
In this paper, we offer an approach that leverages a deceptively simple principle: a set of boundaries \textit{in conjunction with} their medial axes admits a natural way of enumerating points in a given domain. The simplicity of the solution comes directly from three observations regarding the bounding curves of a domain: (1) they automatically induce a medial axis for the domain they bound and (2) they admit a cyclic mapping to the medial axis wherein for each point on the boundary, there exists a straight (\textit{radial}) line to a unique point on the medial axis (the converse is \textit{not} true), and (3) they offer a natural order to sequentially arrange these lines joining the boundary to the medial axis. The implication of these observations is two-fold. First, in Euclidean space where distances are computed by $L_2$ norm, any point inside of the compact domain must be on the line that connects one point on the boundary to another point on the medial axis. Second, the position of the inside point can simply be given by a single parameter, which comes from the linear interpolation of these two endpoints. This gives us a mapping from any point inside the domain to a parameter space.

The polar parameterization is the specific case of this general idea. For any N-dimensional domain bounded by an N-dimensional sphere, the medial axis is a single point, called the center. Any point inside of the domain is simply an interpolation of a boundary point with the center point. 2D is particularly simple, since the boundary curve, i.e. circle can be given a single parameter, which results in $(r,\theta)$ parameterization. One key problem with this purely theoretical solution is the identification of two end-points for any given point inside of the domain. In this paper, we provide a simple discrete approach to find these two points. 

\subsection{Approach}
We begin with a set of Voronoi sites, which we call dipoles, to simultaneously reconstruct an approximate piece-wise linear version of the original boundary and medial axis through Voronoi tessellation. Subsequently, we perform an interpolative re-meshing of this Voronoi tessellation to obtain a partition of the domain that consists of only quadrilaterals and triangles. These quadrilaterals (and triangles) are skinny in the sense that (1) two consecutive vertices are on the boundary and are close to each other, and (2) the other two vertices are on the medial axis and also close to each other. Triangles are formed only when the distance between the two consecutive vertices is zero. 
Using this quad-dominant partition of the domain, it is straightforward to compute any given point in the domain. We first identify the quadrilateral in which the point lies. This translates to computing the bi-linear interpolation.  One parameter in bi-linear interpolation provides the closest positions on the boundary and medial axis. The second parameter provides the parametric position of the given point in terms of the boundary and medial axis positions. 

\begin{figure*}[htbp!]
  \centering
    \begin{subfigure}[t]{0.24\linewidth}
\includegraphics[width=\linewidth]{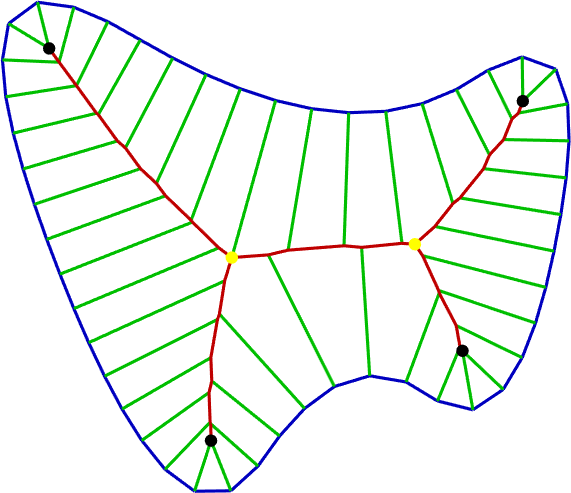}
\caption{Initial Voronoi Decomposition.  }
\label{fig:Figures/00}
    \end{subfigure}
        \hfill
        \begin{subfigure}[t]{0.24\linewidth}
\includegraphics[width=\linewidth]{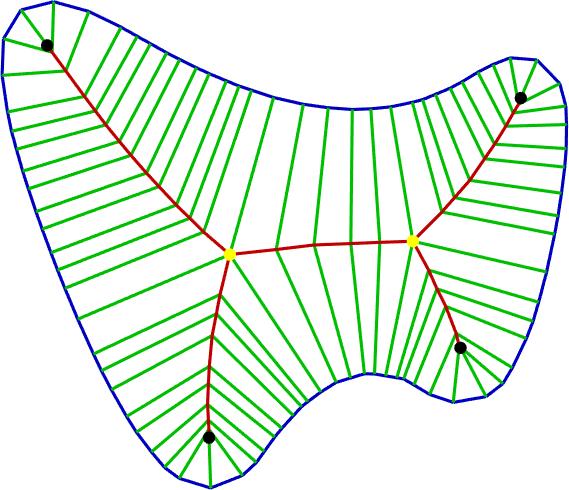}
\caption{Remeshing to create Parametrization.  }
\label{fig:Figures/01}
    \end{subfigure}
        \hfill
            \begin{subfigure}[t]{0.24\linewidth}
\includegraphics[width=\linewidth]{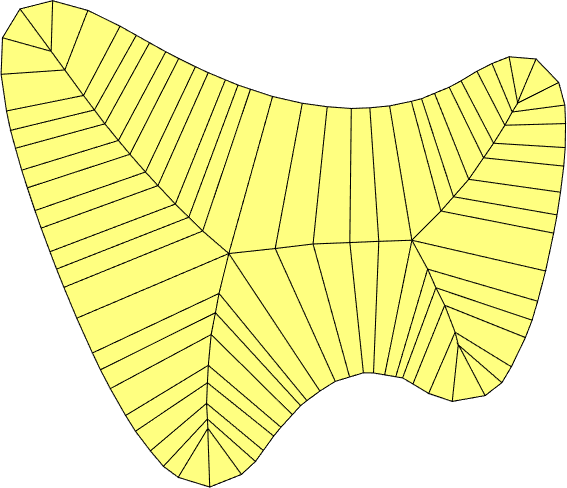}
\caption{Basic Structure.  }
\label{fig:Figures/02}
    \end{subfigure}
        \hfill
    \begin{subfigure}[t]{0.24\linewidth}
\includegraphics[width=\linewidth]{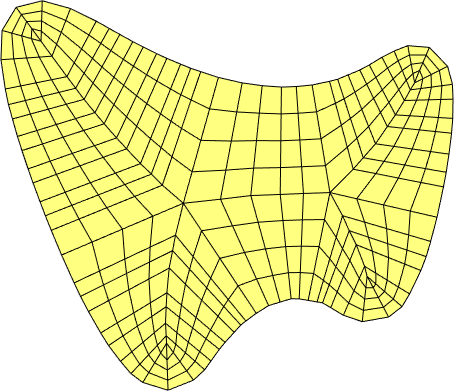}
\caption{Structure showing parametrization.  }
\label{fig:Figures/03}
    \end{subfigure}
        \hfill
    \caption{An example that demonstrates the parameterization of a compact planar domain bounded by a simple (non-self-intersecting) curve. Note that our re-meshing stage is a subdivision algorithm that smooths both the boundary curve and medial axis.  }
    \label{fig:FirstFigure}
 \end{figure*}   

\begin{figure*}[htbp!]
  \centering
    \begin{subfigure}[t]{0.23\linewidth}
\includegraphics[width=\linewidth]{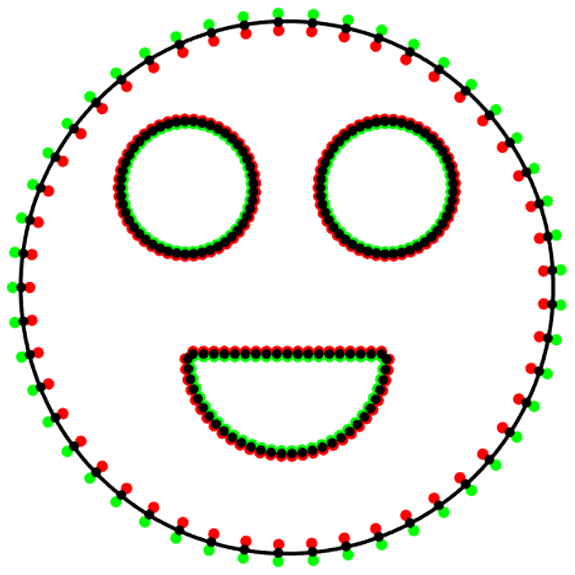}
\caption{Voronoi Sites: Dipoles.  }
\label{fig:Results/Smiley/00}
    \end{subfigure}
        \hfill
    \begin{subfigure}[t]{0.23\linewidth}
\includegraphics[width=\linewidth]{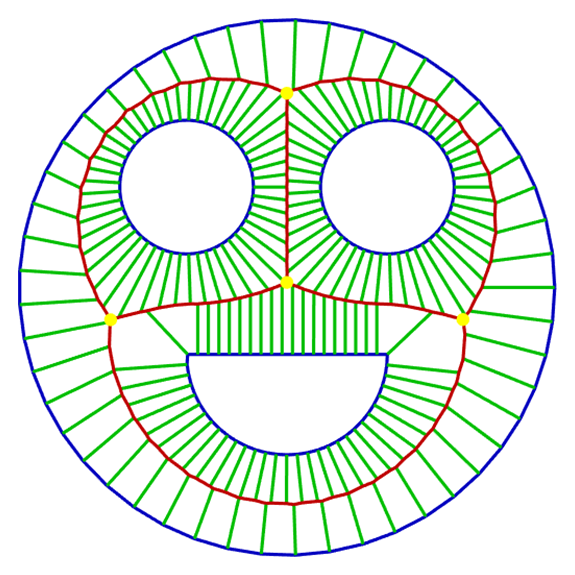}
\caption{Voronoi decomposition.  }
\label{fig:Results/Smiley/01}
    \end{subfigure}
        \hfill
    \begin{subfigure}[t]{0.23\linewidth}
\includegraphics[width=\linewidth]{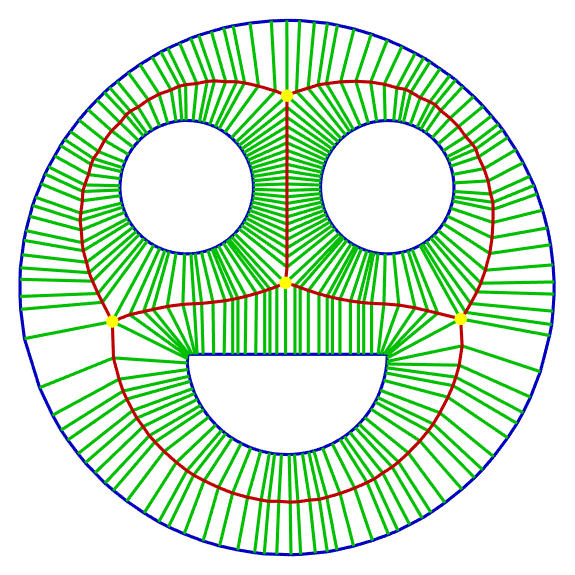}
\caption{Remeshing.  }
\label{fig:Results/Smiley/02}
    \end{subfigure}
        \hfill
    \begin{subfigure}[t]{0.23\linewidth}
\includegraphics[width=\linewidth]{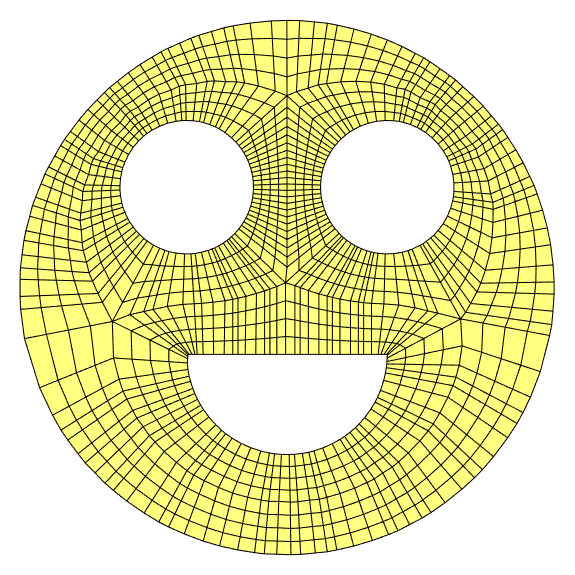}
\caption{Parametrization.  }
\label{fig:Results/Smiley/03}
    \end{subfigure}
        \hfill
    \caption{An example that demonstrates the whole process for a domain with three holes. }
    \label{fig:Smiley}
\end{figure*}

\subsection{Contributions}
We make three main contributions in this paper. First, we introduce a new conceptual framework to parameterize arbitrary compact planar domains by introducing the concept of dipoles. Our framework generalizes the notion of polar coordinates (medial+radial) for arbitrary 2D domains homeomorphic to a planar disk. The framework is further applicable to high-genus as well as multi-component domains. While our approach can potentially work in any dimension, there is a need for a significant amount of additional work to extend it to higher dimensions. Second, we present an algorithm based on Voronoi tessellation to put the theoretical framework into practice for obtaining approximate parametric representations of compact planar domains. For this, we develop a curve-skeleton co-representation using dipoles as Voronoi sites. Using this combined representation, we demonstrate the ability to create a well-defined region using that which needs both a curve as well as a skeleton. Finally, we demonstrate the efficacy of our approach through several examples including a simple method to convert the Voronoi partitions into quadrilateral+triangular meshes that can provide a well-defined parametrization of the domain. This method guarantees to obtain a clean result because of underlying Voronoi tessellation.
\section{Background}
Our work presents an integrated view of several known problems in the geometry and graphics literature such as mean value coordinates, Barycentric coordinates, medial axis computation, and Voronoi tessellations. As such, there is an extensive body of work on each of these topics. Therefore, we specifically focus on closely related previous works on Barycentric coordinates, works on medial axis generation, and works on boundary generation through Voronoi tessellation. 

\subsection{Barycentric Coordinates}

Barycentric coordinates have always been important in Computational Geometry and Computer Graphics because of the difficulties of parameterization of unusual domains. Area-based Barycentric coordinates of triangular shapes that have been widely used in rendering are described by Coxeter in 1969 \cite{coxeter1961introduction}. Barycentric Bilinear coordinates of Quadrilaterals lead the development of tensor product surfaces such as Bezier patches \cite{bartels1995introduction}. For general convex polygons, ex, Wachspress developed a solution using rational polynomials \cite{wachspress1975rational}. During 1990s, until the widespread use \textcolor{blue}{of} subdivision surfaces, an important underlying problem for smooth surface modeling was parameterization of n-sided regular polygonal domains \cite{gregory1990smooth, loop1990generalized,loop1989multisided}. 
Floater developed mean value coordinates for star-shaped generalized polygons \cite{floater2003mean} in 2003. Rustamov identified interior distance Using Barycentric Coordinates \cite{rustamov2009interior}. Ju et al. extended Mean Value Coordinates to Closed Triangular Meshes \cite{ju2005mean}. 

\subsection{Medial Axis Computation}
There has also been a significant amount of work on the construction of the Medial Axis for a given curve using Voronoi diagrams. These methods generally focus on extracting shape information \cite{robinson1992delaunay,ogniewicz1994skeleton}. Since then many papers concentrated on a variety of problems on convergence guarantee \cite{dey2004approximating}, automated skeleton extraction  \cite{sharma2006voronoi}, pruning algorithms \cite{beristain2012pruning, liu2012generation},  continuous skeleton computations \cite{liu2012generation}. 

\subsection{Boundary-Skeleton Co-computation}
In 1998, Amenta showed that Voronoi decomposition can be used for the simultaneous construction of curve and surface boundaries -along with their medial axes- for a given set of points \cite{amenta1998crust,amenta1998new}. The extensions of this approach have been used for integrated skeleton and boundary shape representation for applications such as medical image interpretation \cite{robinson1992integrated}. Faster algorithms are also designed for simultaneous extraction of boundary and medial axis \cite{gold1999crust,gold2001one}

\subsection{Meshing}
Two related topics to our work are quad-dominant remeshing and surface quadrangulation since they can implicitly provide parameterization of surfaces \cite{itoh1998automated,marinov2006robust,alliez2008recent}. Most of these works invoke variational or differential geometric methods such as principal lines of curvature. One interesting work that is closely related to ours is a method for 2D finite element mesh generation by Tam and Armstrong\cite{tam19912d}. This work uses the medial axis to segment the boundary into sub-domains that are subsequently converted into quad-meshes. However, the paper primarily proposes a case-by-case approach for meshing each sub-domain based on the geometric features (concavities, etc.). One of the main advantages of our method is that it is generally applicable to arbitrary domains without any need for explicit sub-domain computation or geometric characterization. 

\subsection{Our work}
While inspired by several prior works on medial axis computation and quad meshing, our goal is not to present new methods for medial axis or quad-dominant meshing. Instead, we specifically focus on the parametrization of compact planar domains. We assume that a domain is bounded by closed, simple (non-intersecting and non-self-intersecting), and oriented curves. For this, we propose a \textit{combined boundary-skeleton representation} of the domain. This representation leverages a deceptively simple yet fundamental insight that the boundary and medial axis, when taken independently, contain only partial information about the domain. However, when combined, they offer a natural way to enumerate points within the domain by allowing us to construct a simple coordinate system inside the domain. Furthermore, this coordinate system directly extends the polar coordinate system for simply-connected domains topologically equivalent to a circular disc. Additionally, for domains with multiple components and holes, we show that it is still possible to construct a similar coordinate system that is composed of multiple single-curve coordinate systems.  Finally, this coordinate system could also be viewed as a type of Barycentric coordinate system formulated as an interpolation between boundary and medial axis points.
\section{Conceptual Preliminaries}

\subsection{Polar Parametrization}
We begin the parametrization of a compact unit circular disc of radius as given by $(r,\theta)$ where $r \in [0,1]$ is the radial direction and $\theta \in [-\pi, \pi]$ is the polar angle or the azimuth direction. The first observation to make here is that $\theta$, which is typically considered as the ``\textit{angular}'' direction is, in effect, parametrizes the arc of the circular boundary of the disc through a linear relation $l = r\theta$, which, for a unit circle is simply $l = \theta$. The second important observation is that\textcolor{blue}{,} this linear relationship corresponds to the other interpretation of the circle: the locus of points equidistant from the pole (center), or, the \textit{iso-distance curve} from the pole. Therefore, the center of the circle is, in effect, the medial axis of the circle as well. These two observations, in conjunction, directly lead to medial parametrization.

\begin{figure}
    \centering
    \includegraphics[width=0.75\linewidth]{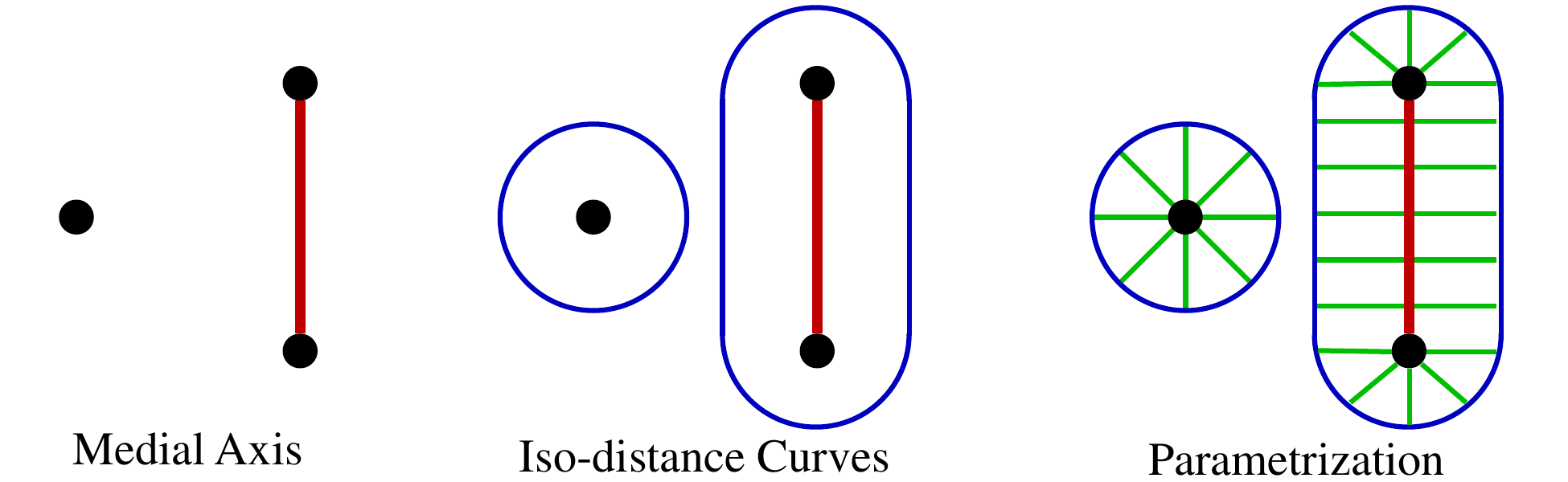}
    \caption{A conceptual example of medial parametrization}
    \label{fig:IL_CONCEPT}
\end{figure}

\subsection{Medial Parametrization}
Consider the elongation of the center of the circle to a straight line (Figure \ref{fig:IL_CONCEPT}). In this case, we can see that the iso-distance offset of the line is a capsule. Conversely, the line is a medial axis to the capsule. Based on this, consider the following two observations:
\begin{enumerate}
    \item \textbf{Radial Direction:} Each point on the boundary maps to a unique point on the medial axis through a straight line. These straight lines essentially generalize the notion of the radial lines (corresponding to the parameter $r$) in the case of the circle. Specifically, the radial lines at the two end points of the medial axis result in the same triangulated (polar) structure of the original circle while generating quadrilateral sub-domains along the curve of the medial axis. 
    \item \textbf{Medial Direction} If the boundary curve (the capsule in our example) is assumed to be oriented and parametrized (e.g. using arc-length parametrization, etc.), then it offers a natural order (or sequence) to the radial lines \textit{around} the medial axis. This is equivalently, the generalization of the angular parameter ($\theta$) for the circle.
\end{enumerate}

Therefore, the combination of the boundary curve (the capsule in our example) combined with the medial axis (the straight line in our example) leads to a natural parametrization of the domain bounded by the boundary curve. Not only that, a closed, simple, and oriented boundary offers a logical sequence of the angular parameter regardless of the topology of the medial axis (such as having a branch). In general, we observe that if a compact planar domain is homeomorphic to a compact unit circular disc, then there exists a mapping between points in the domain to $[0,1]\times[-\pi, \pi]$ domain in polar coordinates. Furthermore, in cases where the domain contains holes and/or multiple components, each bounding curve induces an independent parametrization that covers the domain. In other words, the polar parametrization of the circle is a special case of medial parametrization.

\subsection{Computing Medial Parametrization}
Given the observations above, we note that the so-called medial parametrization is readily computed if both the boundary and the corresponding medial axis are known. For this, we invoke prior work by Sharma et al~\cite{Sharma2009} that establishes the isomorphism between the medial axis and the Voronoi diagram. In other words, given a set of points on a planar boundary curve, the Voronoi tessellation of the plane with those points as the Voronoi sites result in a structure that contains the medial axis. Leveraging this fact, we develop a method to compute medial parametrization.


\section{Algorithm}

\subsection{Overview and Nomenclature}
The primary structure needed for medial parametrization is the combination of boundary and medial curves for a given domain. We begin by observing that typically, a set of points sampled on a curve would result in a Voronoi tessellation that will not consist of the boundary curve itself but only the medial axis. In order to remedy this, we develop a strategy inspired by prior works\cite{turk2002modelling,amenta1998crust} that can simultaneously generate boundary and medial axis using Voronoi tessellations. 
\begin{figure}
    \centering
    \includegraphics[width=0.75\linewidth]{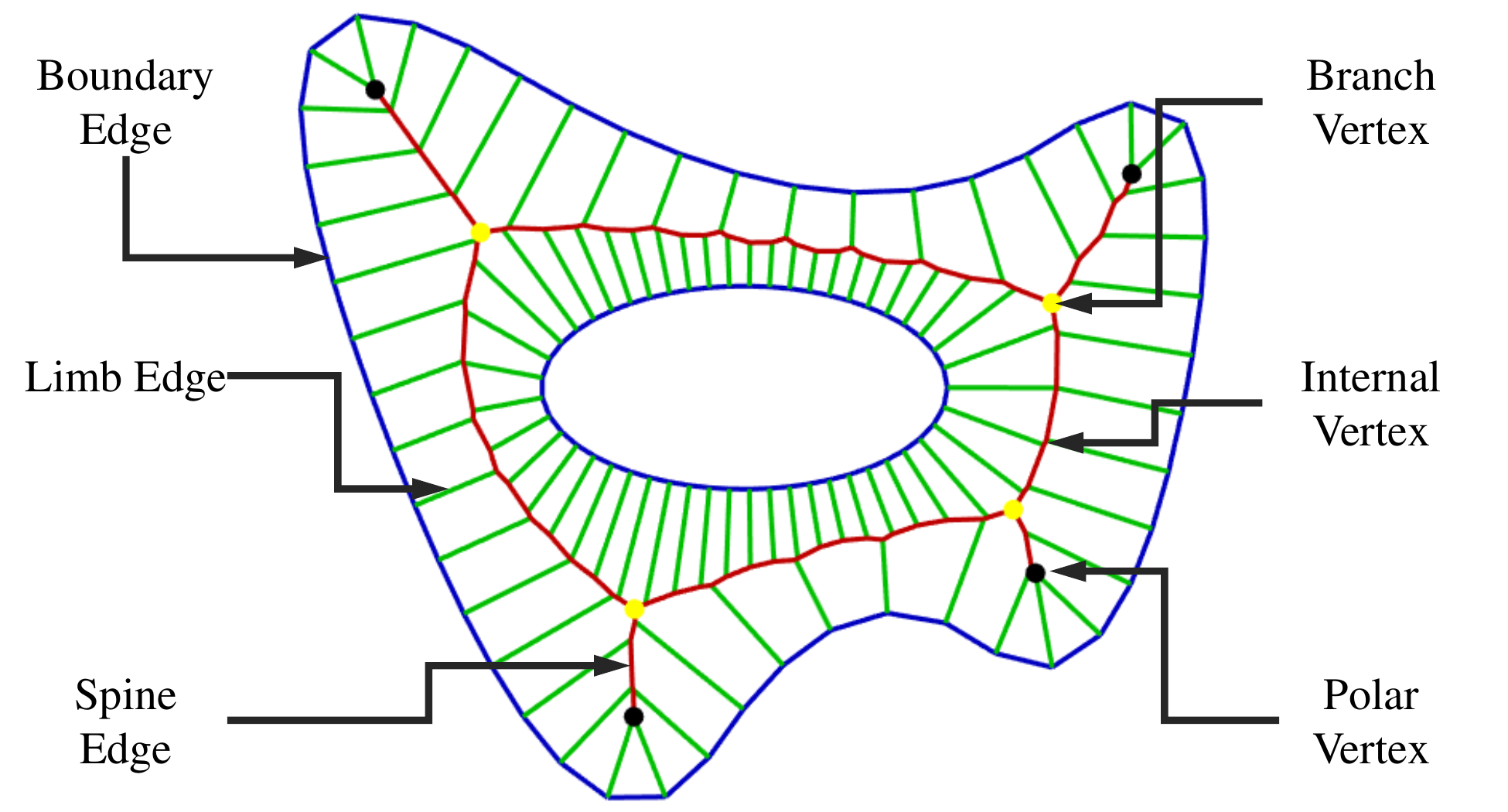}
    \caption{Nomenclature for the medial parametrization algorithm.}
    \label{fig:IL_NOMENCLATURE}
\end{figure}
Our simultaneous construction of boundary and medial axis is inspired by Turk and O'Brien's implicit surface generation \cite{turk2002modelling} wherein they first sample and label points inside and outside the domain.  With this general strategy in mind, we define the following terms for a consistent presentation of our algorithm (Figure \ref{fig:IL_NOMENCLATURE}):

\begin{enumerate}
    \item \textbf{Dipoles:} This refers to a pairs of points $\mathbf{P}^{(+)}_{i}, \mathbf{P}^{(-)}_{i} \in \mathbb{R}^2$ such that $\mathbf{P}^{(+)}_{i}$ is outside and $\mathbf{P}^{(-)}_{i}$ is inside the domain. Furthermore, the line joining these two points is normal to the boundary at point $\mathbf{P}_{i}$.
    \item \textbf{Domain:} This refers to a planar region that is bounded by a set of planar, closed, simple curves $\mathcal{B} = \{B_1,\ldots,B_k\}$ that do not intersect each other. The domain is considered to be compact (i.e. all the curves are included in the domain).
    \item \textbf{Boundary Edges:} These are edges $\mathbf{E}(\mathcal{B}) = \{(i,j) \mid i,j \in [1,|\mathbf{V}(\mathcal{B})|] \}$ on any one of the curves bounding the domain. Here, $\mathbf{V}(\mathcal{B}) = \{\mathbf{P}_{i}^{\mathcal{B}} \in \mathbb{R}^{2}\}$ are the boundary vertices.
    \item \textbf{Spine Edges:} These are edges $\mathbf{E}(S) = \{(i,j) \mid i,j \in [1,|\mathbf{V}(\mathcal{S})|] \}$ on the medial axis of the domain.
    \item \textbf{Limb Edges:} These are edges $\mathbf{E}(L) = \{(i,j) \mid i,j \in [1,|\mathbf{V}(\mathcal{L})|] \}$ that connect each point on the boundary to point on the spine. Note that no additional vertices are required to create the limb.
    \item \textbf{Spine Vertices:} These are vertices $\mathbf{V}(\mathcal{S}) = \{\mathbf{P}_{i}^{\mathcal{S}} \in \mathbb{R}^{2}\}$ on the medial axis of the domain. There are three types of spine vertices, namely:
    \begin{enumerate}
        \item \textit{Interior Vertices:} Vertices with valency $2$.
        \item \textit{Branch Vertices:} Vertices with valency $ > 2$.
        \item \textit{Polar Vertices:} Vertices with valency $1$.
    \end{enumerate}
    \textbf{[Remark:]} These valencies include exclusively the spine edges (i.e. edges that compose the medial axis and \textit{not} the limb edges.
\end{enumerate}

Our algorithm has two broad steps: (1) simultaneous co-construction of boundary and medial axis using Voronoi tessellation with dipoles, and (2) interpolative re-meshing of the Voronoi tessellation. The second step specifically aims at generating a \textit{boundary---limb---spine} sequence such that all re-generated internal and branch vertices on the spine are 4-valent while the polar vertices on the spine have valency $\geq 4$. In other words, all faces of the re-meshed Voronoi tessellation incident on interior and branch vertices on the spine are quadrilaterals while the faces containing polar vertices are triangles.

As such, our algorithm is agnostic to the number of components and holes in the domain. However, for simplicity of description, we will assume that the domain is bounded by only one boundary curve (i.e. it is a single component without holes). 

\begin{figure}[htbp!]
\centering
      \begin{subfigure}[t]{0.40\linewidth}
\includegraphics[width=\linewidth]{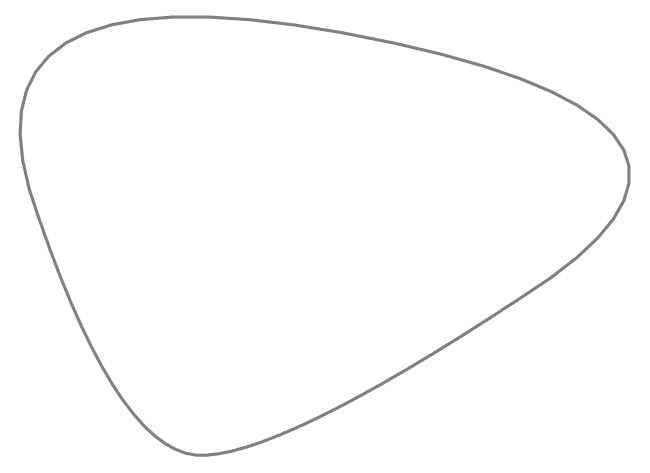}
\caption{Initial Curve.  }
\label{fig:Results/Delta/00}
    \end{subfigure}
    \begin{subfigure}[t]{0.40\linewidth}
\includegraphics[width=\linewidth]{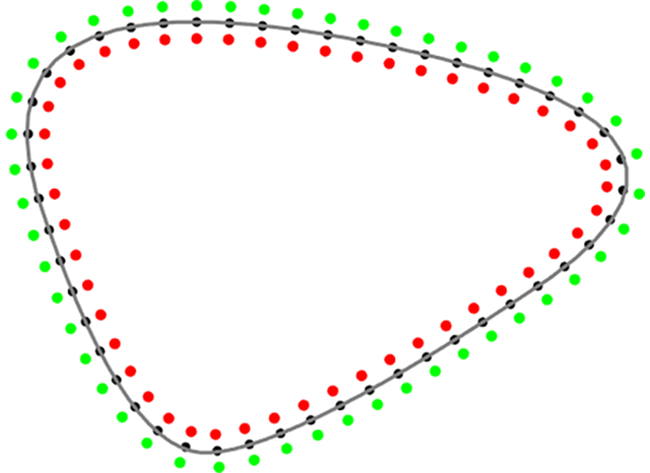}
\caption{Dipoles.  }
\label{fig:Results/Delta/01}
    \end{subfigure}
        \hfill
    \caption{An example demonstrating dipole creation. }
    \label{fig:algoDipole}
\end{figure}

\subsection{Step 1: Dipole Creation}

We first start with a $G^1$ continuous curve\footnote{The dipoles can also be used with $G^0$ curves, i.e. polygons. For polygons, we need to keep sharp corners, which requires a little more care. To simplify the exposure in this paper, we focus on $G^1$ continuous curves.}. We begin by sampling $n$ points on a given curve. For multi-component domains and domains with holes, each curve is individually sampled. Note that the distance between the samples does not have to be the same for this method to work. Let $\mathbf{P}_{i}$ and $\vec{T}_{i}$ denote the position and tangent vector of the $i^{th}$ sample point where $i \in [1,\ldots,N]$. The tangent vector is well-defined everywhere since we start with $G^1$ continuous curve. We then create a pair of Voronoi sites in each sample $i$. 

\begin{figure}[htbp!]
  \centering
      \begin{subfigure}[t]{0.23\linewidth}
\includegraphics[width=\linewidth]{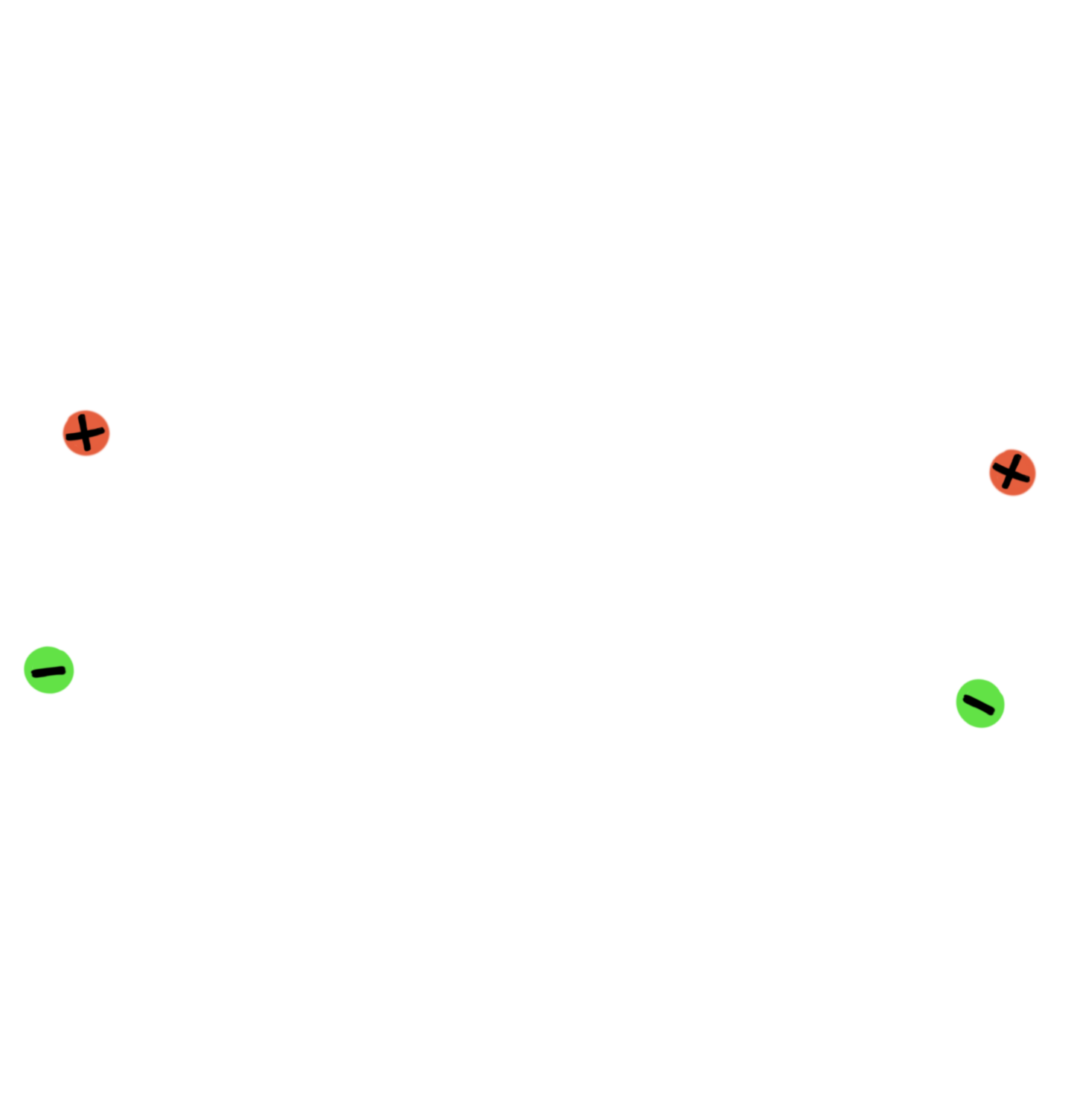}
\caption{Two sets of dipoles.  }
\label{fig:Figures/Dipoles0}
    \end{subfigure}
        \hfill
        \begin{subfigure}[t]{0.16\linewidth}
\includegraphics[width=\linewidth]{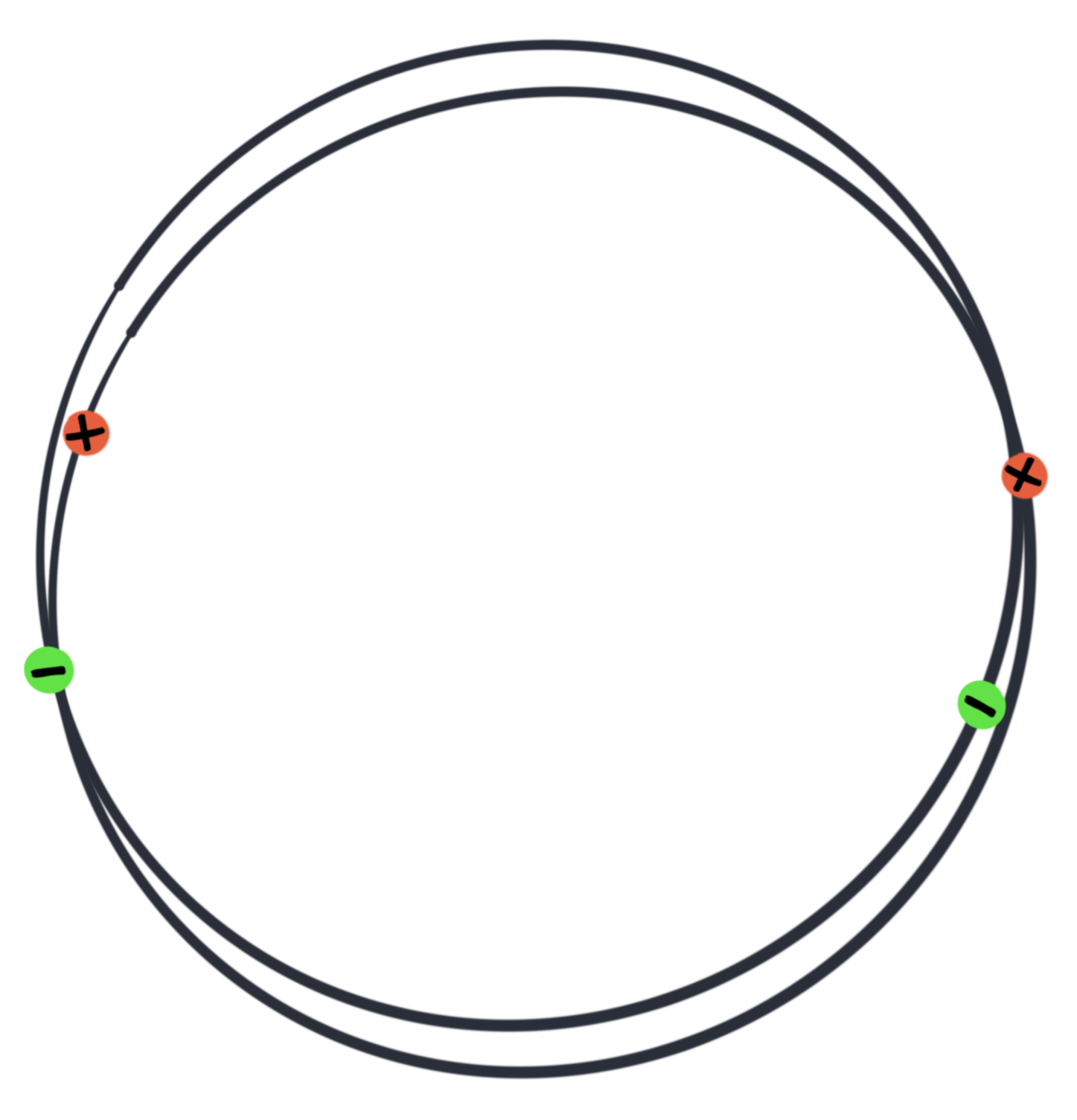}
\caption{Not one but two circles.  }
\label{fig:Figures/Dipoles1}
    \end{subfigure}
        \hfill
    \begin{subfigure}[t]{0.23\linewidth}
\includegraphics[width=\linewidth]{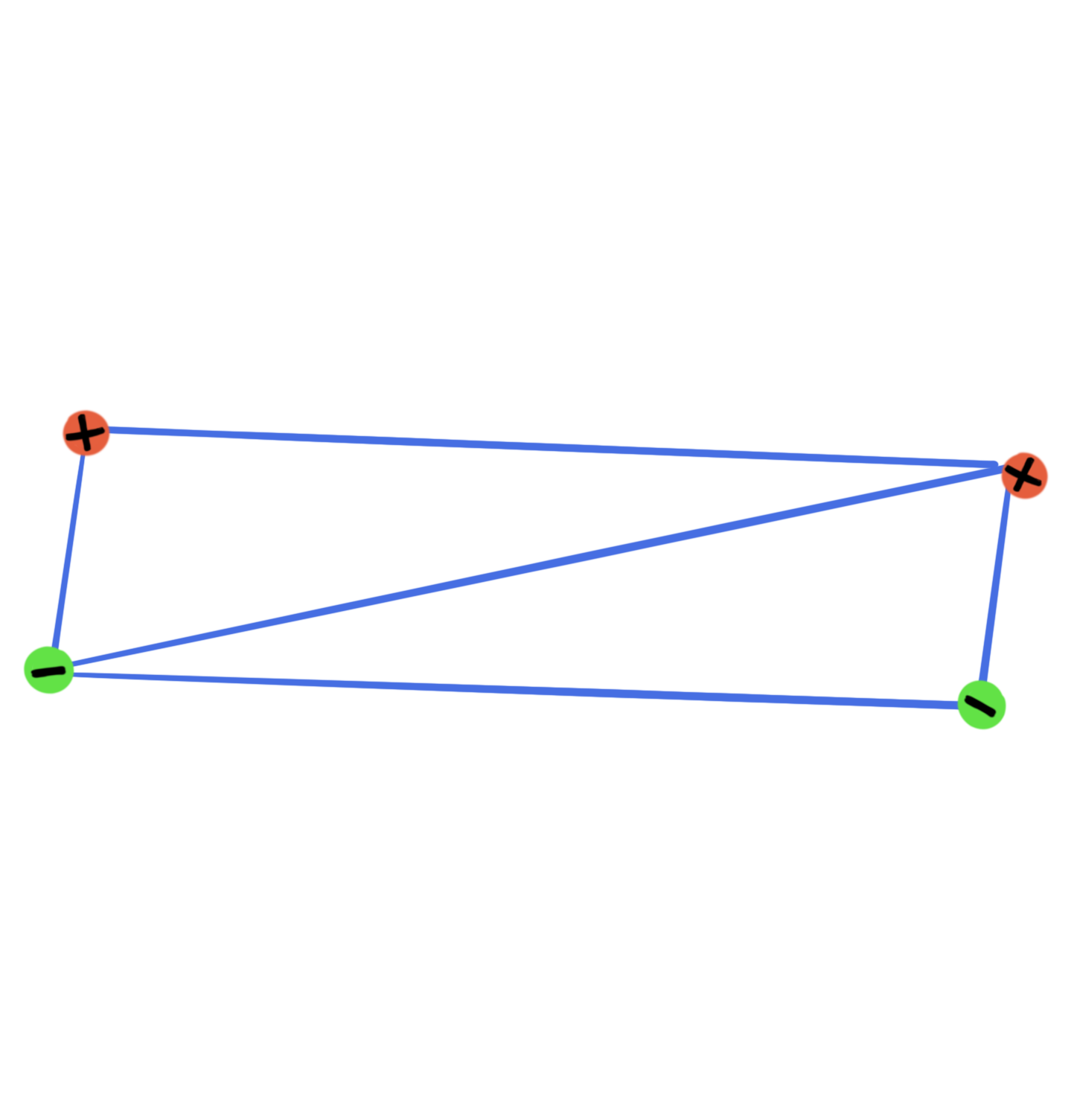}
\caption{Delaunay triangulation.  }
\label{fig:Figures/Dipoles2}
    \end{subfigure}
        \hfill
    \begin{subfigure}[t]{0.23\linewidth}
\includegraphics[width=\linewidth]{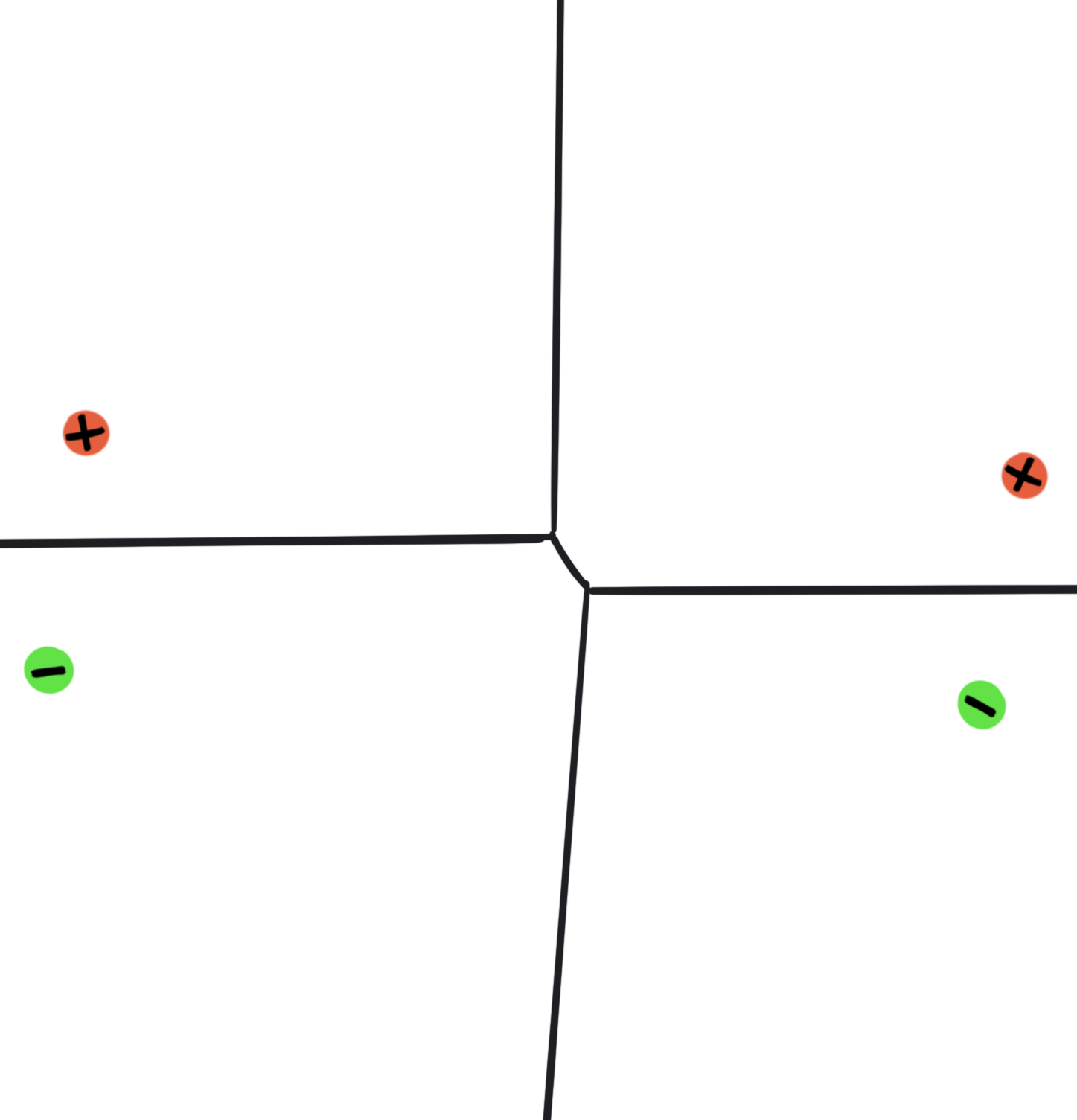}
\caption{Voronoi decomposition.  }
\label{fig:Figures/Dipoles3}
    \end{subfigure}
        \hfill
    \caption{An example demonstrating that dipoles that are away from each other can create two 3-valent vertices that are very close to each other. Even if we put these four points in the same circle, we sometimes face this situation because of numerical error. Therefore, in our experience, it is sufficient to keep them apart from each other. Of course, there should be no other part of the curve that is close to this region.  }
    \label{fig:Dipoles}
\end{figure}

\subsubsection{Dipole Generation \& Voronoi Tessellation}
Let $\mathbf{P}^{(+)}_{i}$ and $\mathbf{P}^{(-)}_{i}$ denote these two dipole points outside and inside of the domain respectively.  These two points are created based on the following three conditions: 
\begin{eqnarray}
|\mathbf{P}^{(+)}_{i} - \mathbf{P}_{i}| &=& |\mathbf{P}^{(-)}_{i} - \mathbf{P}_{i}|  \label{cond1}\\
\vec{T}_{i} \bullet (\mathbf{P}^{(+)}_{i} - \mathbf{P}^{(-)}_{i}) &=& 0 \label{cond2} \\
2|\mathbf{P}^{(+)}_{i} - \mathbf{P}_{i}| &<<& |\mathbf{P}_{i+1} - \mathbf{P}_{i-1}| \label{cond3}
 \end{eqnarray}
where $\bullet$ is the dot product. 
We simply populate the boundary with these dipoles and compute the Voronoi tessellation with the dipole points as the Voronoi sites. The resulting Voronoi tessellation results in both an approximation of the boundary and medial axis (Figure~\ref{fig:Results/Delta/02}). 
Condition~\ref{cond1} guarantees that one of the lines in Voronoi decomposition goes through $\mathbf{P}_{i}$. Condition~\ref{cond2} guarantees that this particular line is tangent to the boundary curve at $\mathbf{P}_{i}$. For the best results, the last condition must normally be ``\textit{two consecutive dipoles must be in the same circle (i.e. they should form a cyclic quadrilateral)}'' to obtain 4-valence vertices in the boundary. Condition~\ref{cond3} is really a compromise to simplify the algorithm. If two dipoles are reasonably far away from each other, they still form a \textit{near-cyclic} quadrilateral (See Figure~\ref{fig:Figures/Dipoles1}). As a result, they tend to create two three valence Voronoi vertices that are connected by a very short edge (See Figure~\ref{fig:Figures/Dipoles3}). While this operation changes position and tangent in the sample point, the change is practically negligible in our experience. In the final step, we collapse edges with a length less than a pre-determined threshold, say $\varepsilon$.

\subsubsection{Boundary and Medial axis Identification}
Let $\mathcal{T}(\mathbf{V},\mathbf{F})$ denote the Voronoi tessellation. Note that each Voronoi cell $f = \{i_1, i_2,\ldots,i_k\} \in \textbf{F}$ corresponds to a unique site in the set of dipoles $\{\mathbf{P}^{(+)}_{i}\} \cup \{\mathbf{P}^{(-)}_{i}\}$ where each site is labeled either as a point inside ($\mathbf{P}^{(+)}_{i}$) or outside ($\mathbf{P}^{(-)}_{i}$) the domain. Using this one-to-one site-cell mapping, we simply apply the dipole labels to each of the Voronoi cells. This is readily allows us to classify each edge $(i,j) \in \mathbf{E}(\mathcal{T})$ of the Voronoi tessellation as follows:
\begin{itemize}
    \item If $\mathbf{Q}_{i}, \mathbf{Q}_{j} \in \{\mathbf{P}^{(-)}_{i}\}$, then $label(i,j)~=~In$ 
    \item Else If $\mathbf{Q}_{i}, \mathbf{Q}_{j} \in \{\mathbf{P}^{(+)}_{i}\}$, then $label(i,j)~=~Out$
    \item Else append $(i,j)$ to $\mathbf{E}(\mathcal{B})$
\end{itemize}
The edges labeled $In$ comprise both the limb and spine edges while those labeled $On$ are the boundary edges. Spine edges (i.e. edges on the medial axis) are those that do not have a vertex on the boundary. In order to extract the medial axis, we classify all interior edges as follows:

\begin{itemize}
    \item If $label(i, j) = In$ \& $\exists (a, b) \in \mathbf{E}(\mathcal{B}) \mid i \in \{a,b\}$ or $j \in \{a,b\}$, then: append $(i,j)$ to $\mathbf{E}(\mathcal{L})$
    \item Else append $(i,j)$ to $\mathbf{E}(\mathcal{S})$
\end{itemize}

Once we have the boundary, spine, and limb edges labeled in the Voronoi tessellation, we have the complete structure of the domain.

\subsection{Step 2: Interpolative Re-meshing of Voronoi Tessellation}
One of the main characteristics of the Voronoi tessellation resulting from the dipoles is that most cells are non-quadrilateral and most vertices are 3-valence vertices. This is not just true for the dipoles discussed above, but also for arbitrary arrangement of Voronoi sites. Specifically, if we consider the spine (i.e. the medial axis given by the Voronoi tessellation), we typically obtain 3-valence vertices on the spine \textit{including the limb edges}. While we did not find a comprehensive theoretical or experimental treatise on the valency and cell statistics for Voronoi tessellations, we believe that getting quad-cells and 4-valence vertices are low-likelihood events for Voronoi tessellations for a random set of sites. For instance, a rectangular and trapezoidal grid of sites are the only two site configurations that we currently know to produce a Voronoi diagram consisting of purely quadrilateral. 
\begin{figure}[htbp!]
  \centering
  \centering
      \begin{subfigure}[t]{0.45\linewidth}
\includegraphics[width=\linewidth]{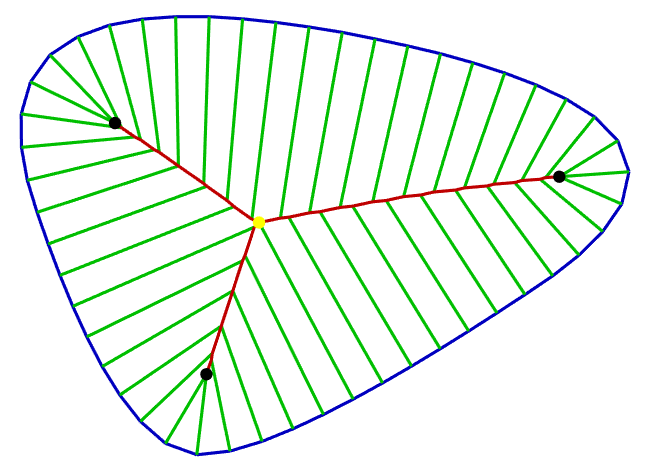}
\caption{Initial Voronoi decomposition.  }
\label{fig:Results/Delta/02}
    \end{subfigure}
        \hfill
    \begin{subfigure}[t]{0.45\linewidth}
\includegraphics[width=\linewidth]{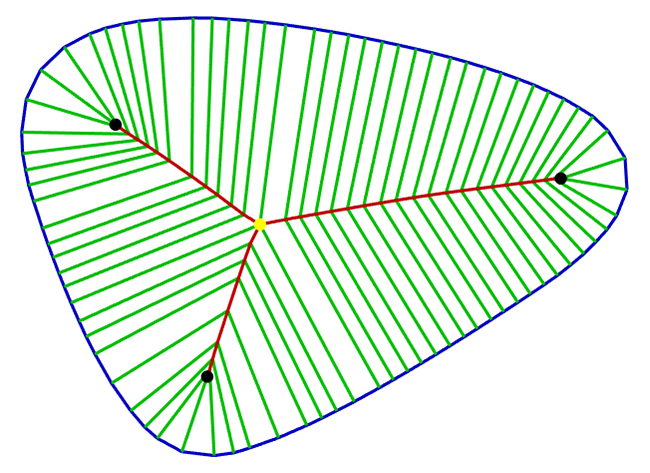}
\caption{Result of remeshing.  }
\label{fig:Results/Delta/03}
    \end{subfigure}
        \hfill
    \caption{An example demonstrating the remeshing process. Remeshing that turns the original Voronoi decomposition that consists of mostly 3-valence vertices and non-quadrilateral polygons into a structure that consists of mostly 4-valence vertices and quadrilaterals. This process also smooths the original zigzag patterns that are caused by Voronoi subdivision. }
    \label{fig:algorithmRemeshing}
\end{figure}
\begin{figure*}
    \centering
    \includegraphics[width=1.0\linewidth]{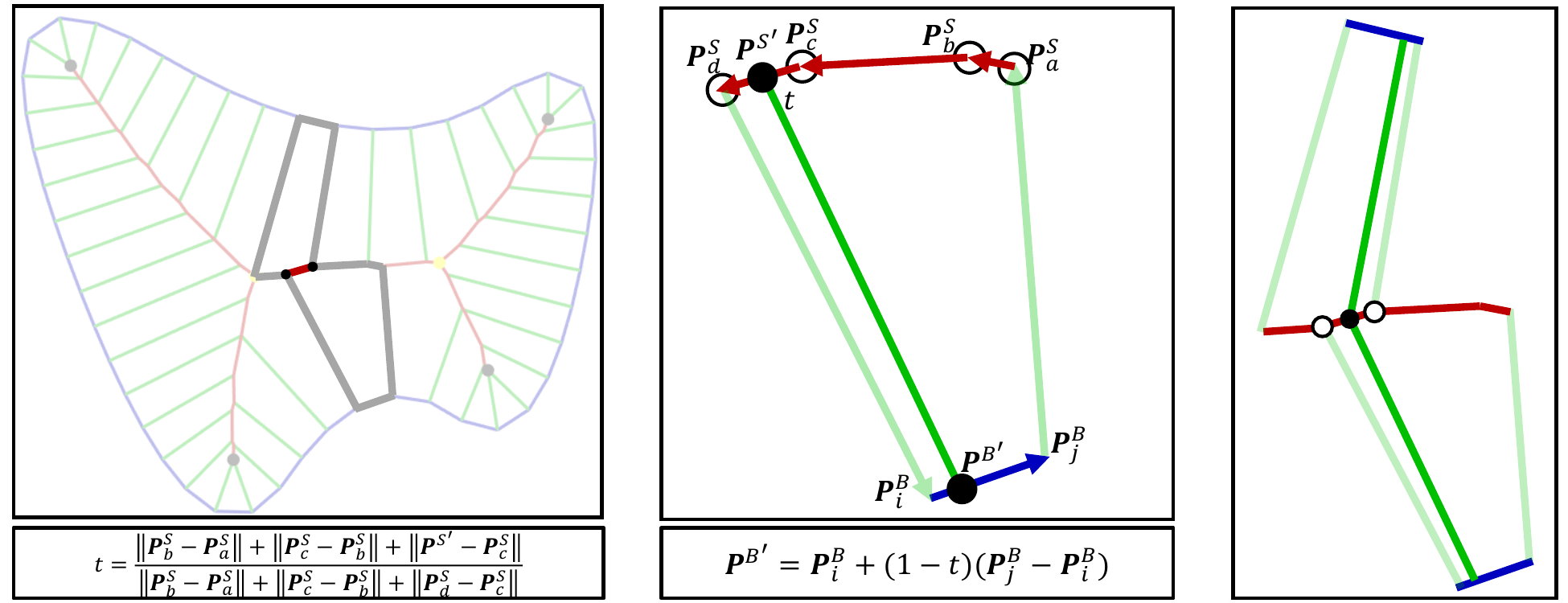}
    \caption{Interpolative re-meshing of the Voronoi tessellation is illustrated. For each spine edge (shown in \textcolor{myR}{red} in the left), we first identify the unique pair of faces containing this edge. In a given face in this face-pair (middle panel), we first compute the mid-point of the spine edge in question and subsequently compute the parameter $t$ based on the ratio of the arc-length from the starting point $\mathbf{P}^{\mathcal{S^{'}}}_{a}$ of the spine edge segment to the mid-point ($\mathbf{P}^{\mathcal{S^{'}}}$) to the total arc-length of the spine edge segment (the \textcolor{myR}{red} segment in middle panel). Finally, we use $t$ to compute a new point on the boundary (shown in \textcolor{myB}{blue}). Note that for an oriented polygon, the parameter is changed to $1-t$. Finally, the new interpolated limb (shown in \textcolor{myG}{green}) is obtained by creating an edge between the mid-point on the spine ($\mathbf{P}^{\mathcal{S^{'}}}$) and the interpolated boundary point ($\mathbf{P}^{\mathcal{B^{'}}}$). The result is a pair of new limbs (right panel)}
    \label{fig:IL_INTERPOLATION}
\end{figure*}

Our main goal is to re-mesh the initial Voronoi tessellation so as to create quadrilateral cells between the boundary and the spine of the domain \textit{with the exception of the polar vertices}, where the faces will be triangular. A direct implication of this goal is that each spine vertex in the re-meshed Voronoi tessellation will have a valency of 4 \textit{except for polar vertices} (see Figure~\ref{fig:algorithmRemeshing} for an example). This gives us a direct method (Figure \ref{fig:IL_INTERPOLATION}) to generate a new set of limbs as follows:

\begin{enumerate}
    \item Initialize new limb edges as: $\mathbf{E}(\mathcal{L}^{'})=\emptyset$
    \item For each spine edge $(i,j) \in \mathbf{E}(\mathcal{S})$
    \begin{enumerate}
        \item Find faces $f_1, f_2 \in \mathbf{F}(\mathcal{T}) \mid (c, d) \in f_1, f_2$
        \item Suppose $f_1 = [i,j,a,b,c,d,\ldots]$ where $i$ and $j$ are boundary vertices, find the chain of spine edges $[(a,b),(b,c),(c,d),\ldots$
        \item Compute new spine point as the mid-point $\mathbf{P}^{\mathcal{S}^{'}} = 0.5(\mathbf{P}^{\mathcal{S}}_{c} + \mathbf{P}^{\mathcal{S}}_{d})$
        \item Compute the total arc-length of the spine edge chain as: $||s_{(a,b,c,\ldots)} = \mathbf{P}^{\mathcal{S}}_{a} - \mathbf{P}^{\mathcal{S}}_{b}|| + \ldots$
        \item Compute the arc length of the spine edge chain up to $(c,d)$ as given by: $s_{a,b,c,d}||\mathbf{P}^{\mathcal{S}}_{a} - \mathbf{P}^{\mathcal{S}}_{b}|| + \ldots + 0.5||\mathbf{P}^{\mathcal{S}}_{c} - \mathbf{P}^{\mathcal{S}}_{d}||$
        \item Determine the arc-length parameter as $t=\frac{s_{(a,b,c,d)}}{s_{(a,b,c,\ldots)}}$
        \item Interpolate new boundary point $\mathbf{P}^{\mathcal{B}^{'}} = \mathbf{P}^{\mathcal{B}}_{i} + (1-t)(\mathbf{P}^{\mathcal{B}}_{j}-\mathbf{P}^{\mathcal{B}}_{i}) $.
        \item Create a new limb in $f_1$ between $\mathbf{P}^{\mathcal{B}^{'}}$ and $\mathbf{P}^{\mathcal{S}^{'}}$
        \item Repeat the process for $f_2$ to obtain the second limb.
        \item Append new limb-pair to $\mathbf{E}(\mathcal{L}^{'})$
    \end{enumerate}
    \item Order the new indexed vertices $\mathbf{P}^{\mathcal{B}^{'}}_{i} \in \mathcal{B}^{'}$ in a sequence.
    \item Orient $\mathcal{B}^{'}$ such that it has a positive area (i.e. it is oriented counter-clockwise in the plane).
    \item Order each limb edge in $\mathcal{L}^{'}$ according to the order induced by $\mathcal{B}^{'}$
\end{enumerate}

Once completed, the steps above provide a trivial way to construct faces for the new mesh. One need \textcolor{blue}{to}just traverse through the boundary list $\mathcal{B}^{'}$, determine the two connected limbs in $\mathcal{L}^{'}$ (which are also sequenced) and connect the spine vertices to construct a quadrilateral (or a triangle at the polar vertices). Another interesting observation to make here is that this process of interpolating at the mid-point of the spine edges effectively performs simultaneous corner-cutting on the medial axis and the boundary thereby smoothing the original zigzag patterns that are caused by Voronoi subdivision.
\begin{figure*}[htbp!]
  \centering
    \begin{subfigure}[t]{0.24\linewidth}
\includegraphics[width=\linewidth]{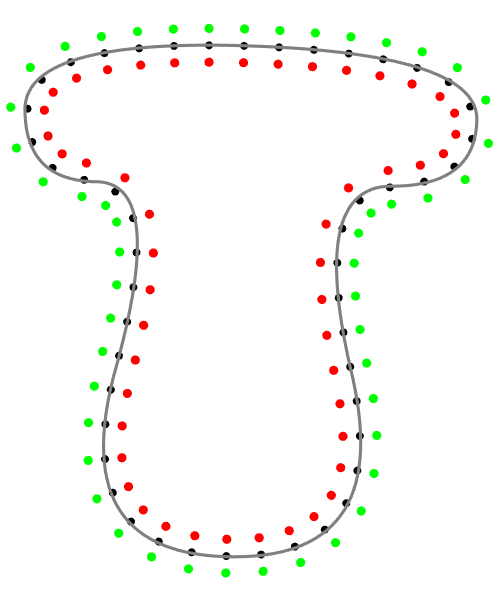}
\caption{Dipoles.  }
\label{fig:Results/Hammer/01}
    \end{subfigure}
        \hfill
    \begin{subfigure}[t]{0.24\linewidth}
\includegraphics[width=\linewidth]{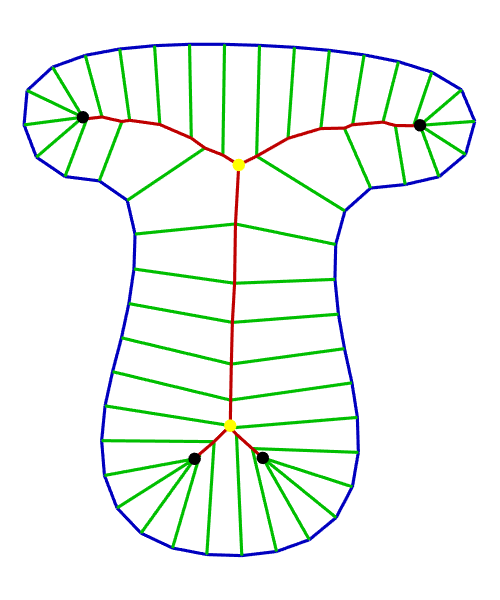}
\caption{Voronoi decomposition.  }
\label{fig:Results/Hammer/02}
    \end{subfigure}
        \hfill
    \begin{subfigure}[t]{0.24\linewidth}
\includegraphics[width=\linewidth]{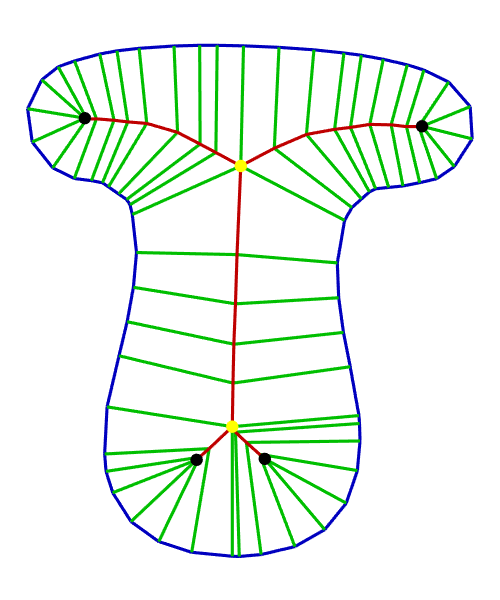}
\caption{Remeshing.  }
\label{fig:Results/Hammer/03}
    \end{subfigure}
        \hfill
    \begin{subfigure}[t]{0.24\linewidth}
\includegraphics[width=\linewidth]{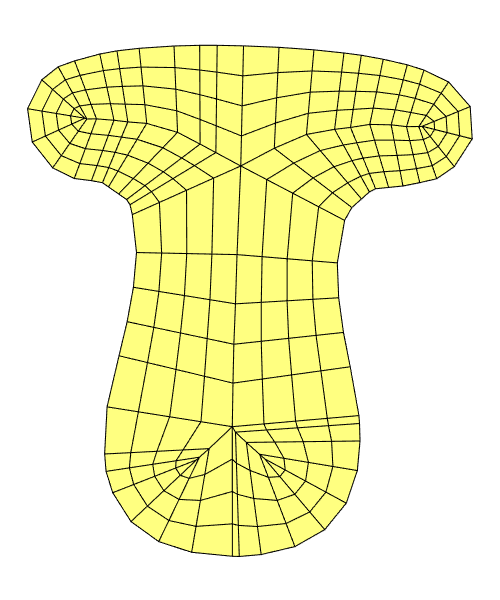}
\caption{Parametrization.  }
\label{fig:Results/Hammer/04}
    \end{subfigure}
        \hfill
    \caption{The process of parametrization of a star shape. Note that even such a simple shape can have a medial axis that is more complicated than expected such as the second branch at the left. }
    \label{fig:Hammer}
\end{figure*}

\begin{figure*}[t]{
\centering
\includegraphics[width=1\textwidth]{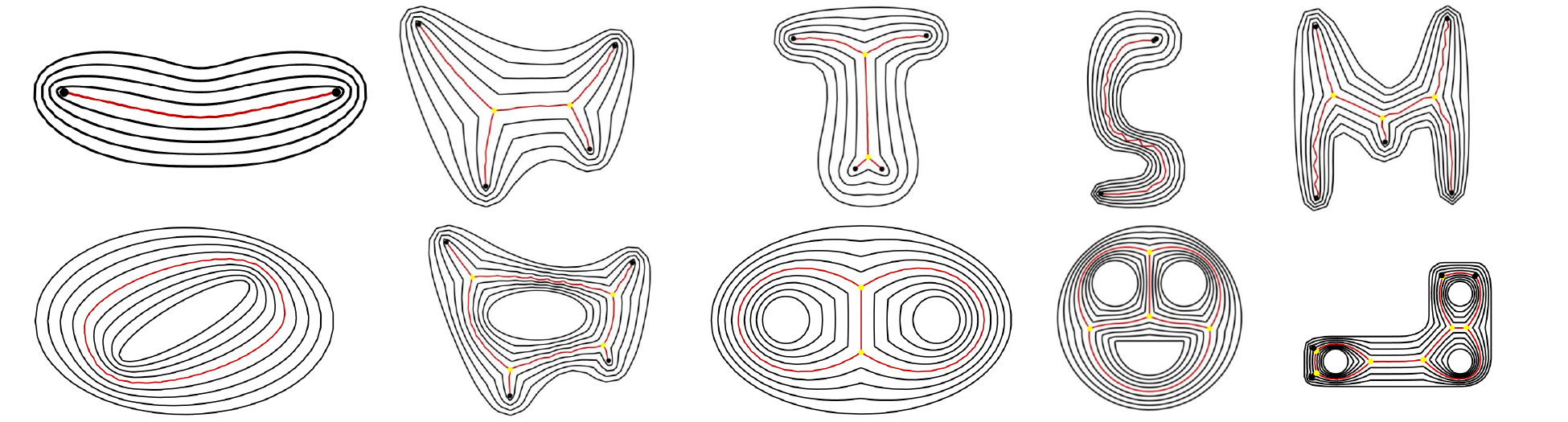}
\caption{Examples demonstrating contours generated from parametrization of the closed domains.\textcolor{myR}{Red} represents the spine edges. Row(above) represents a planar domain bounded by a set of single closed curves and Row(Bottom) represents a planar domain bounded by a set of multiple closed curves}
\label{fig:RE_CONTOUR}
}
\end{figure*}

\begin{figure*}[t]{
\centering
\includegraphics[width=1\textwidth]{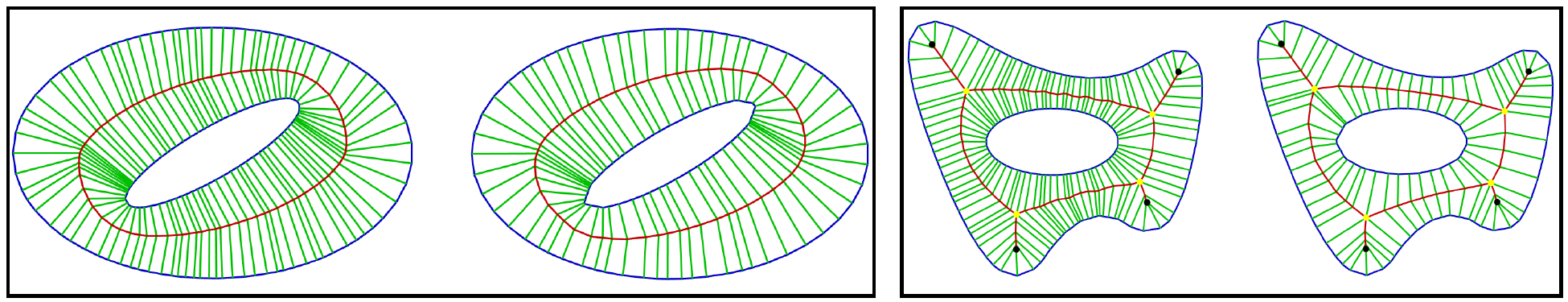}
\caption{Examples (right in each box) demonstrating input curve length dependent sampling for Voronoi decomposition of the input domain and (left in each box) represents equidistant sampling using same number of sample points for Voronoi decomposition of the input domain.}
\label{fig:RE_SAMPLING}
}
\end{figure*}

\section{Experiments and Results}

\subsection{Overview}
\begin{figure}[htbp!]
  \centering
    \begin{subfigure}[t]{0.20\linewidth}
\includegraphics[width=\linewidth]{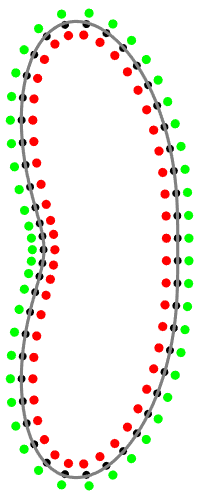}
\caption{Voronoi Sites: Dipoles.  }
\label{fig:Results/Bean/01}
    \end{subfigure}
        \hfill
    \begin{subfigure}[t]{0.20\linewidth}
\includegraphics[width=\linewidth]{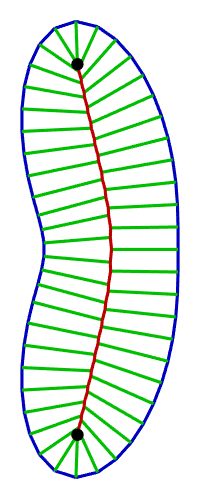}
\caption{Voronoi decomposition.  }
\label{fig:Results/Bean/02}
    \end{subfigure}
        \hfill
    \begin{subfigure}[t]{0.20\linewidth}
\includegraphics[width=\linewidth]{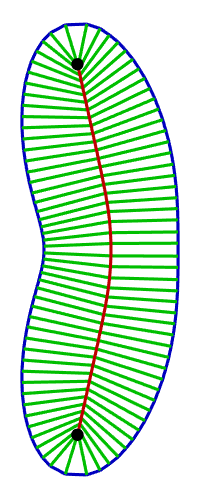}
\caption{Remeshing.  }
\label{fig:Results/Bean/03}
    \end{subfigure}
        \hfill
    \begin{subfigure}[t]{0.20\linewidth}
\includegraphics[width=\linewidth]{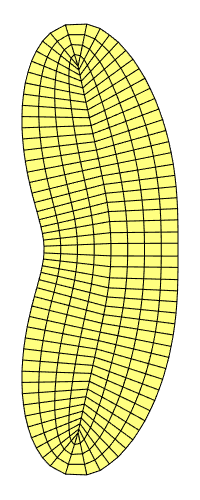}
\caption{Parametrization.  }
\label{fig:Results/Bean/04}
    \end{subfigure}
        \hfill
    \caption{The process of parameterization of a bean shape. }
    \label{fig:Bean}
\end{figure}

In this paper, we perform a number of experiments to investigate the efficacy of our methodology. We conducted our experiments with compact planar domains generated by two categories of boundary curves (1) single closed curve and (2) multiple closed curves. In this paper, we have experimented with $G^1$ continuous curves. Voronoi decomposition of a single closed curve as shown in Figure~\ref{fig:Results/Bean/02} shows that the shape has a continuous medial axis. While another experiment with Figure~\ref{fig:Figures/01} represents a medial axis with branches. Even for simple shapes (Figure \ref{fig:Hammer}), the medial axis can have unexpected branches. 

Figures~\ref{fig:StarwithHole} and ~\ref{fig:StarwithEllipticHole} extend our exploration into multi-curve domain. Figure~\ref{fig:Results/Ellipse10degreeHole/00},~\ref{fig:Results/Ellipse20degreeHole/00} and~\ref{fig:Results/Ellipse30degreeHole/00} represent closed domain shapes with single hole. Domains with elliptical holes arranged in an ascending order of angles simulates the effect of twisting holes on parametrization. We further explore the effect of multiple holes as shown in Figures~\ref{fig:Smiley}and ~\ref{fig:Ellipse2Holes}. These explorations give an insight about the parametrization of different compact planar domain.

\subsection{Key Observations}
In order to study medial parametrization, we study (1) the meshing induced by it and (2) the iso-parametric contours along the radial direction (Figure \ref{fig:RE_CONTOUR}). We further make observations regarding the effect of different curve sampling schemes on the resulting medial axis and the subsequent parametrization (Figure \ref{fig:RE_SAMPLING}). Specifically, we followed two strategies: (1) an equal number of samples on each curve of a domain and (2) length-dependent sampling wherein we set a sample size for the longest curve and sample the others such that the edge length after re-sampling is equal to that on the longest curve.

\begin{figure*}[htbp!]

  \centering
      \begin{subfigure}[t]{0.24\linewidth}
\includegraphics[width=\linewidth]{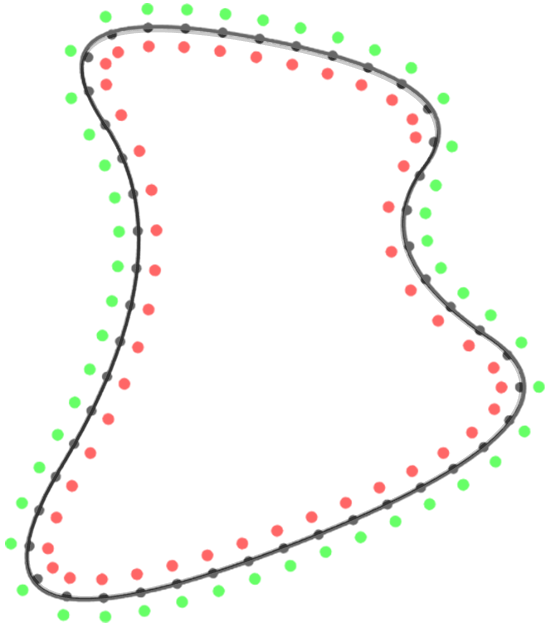}
\caption{Voronoi Sites: Dipoles. }
\label{fig:Results/Star/00}
    \end{subfigure}
        \hfill
    \begin{subfigure}[t]{0.24\linewidth}
\includegraphics[width=\linewidth]{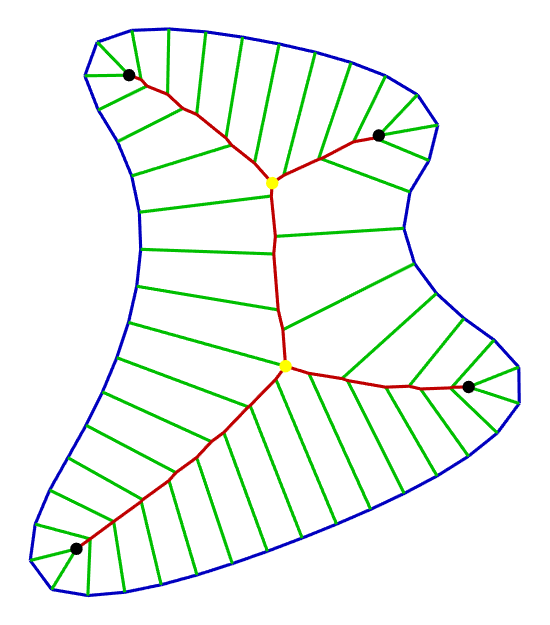}
\caption{Voronoi decomposition.   }
\label{fig:Results/Star/01}
    \end{subfigure}
        \hfill
    \begin{subfigure}[t]{0.24\linewidth}
\includegraphics[width=\linewidth]{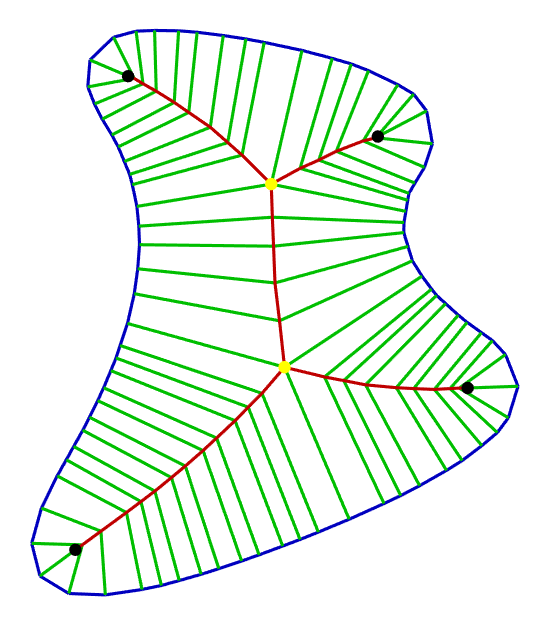}
\caption{Remeshing.     }
\label{fig:Results/Star/02}
    \end{subfigure}
        \hfill
    \begin{subfigure}[t]{0.24\linewidth}
\includegraphics[width=\linewidth]{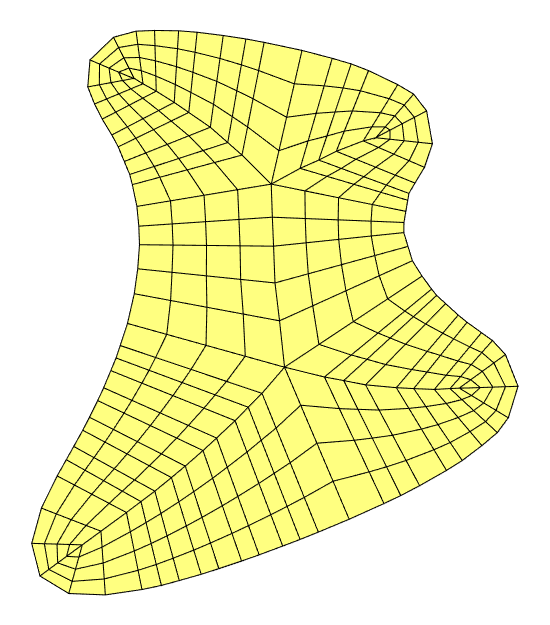}
\caption{Parametrization. }
\label{fig:Results/Star/03}
    \end{subfigure}
        \hfill
    \caption{The process of parameterization of a star-shaped domain. }
    \label{fig:Star}

  \centering
      \begin{subfigure}[t]{0.24\linewidth}
\includegraphics[width=\linewidth]{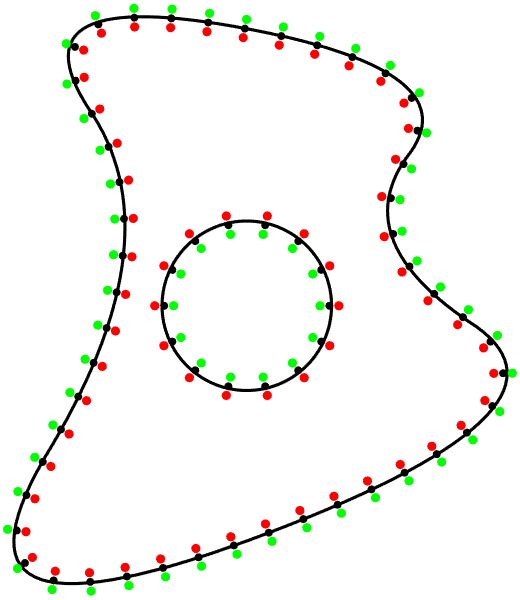}
\caption{Voronoi Sites: Dipoles. }
\label{fig:Results/StarwithHole/00}
    \end{subfigure}
        \hfill
    \begin{subfigure}[t]{0.24\linewidth}
\includegraphics[width=\linewidth]{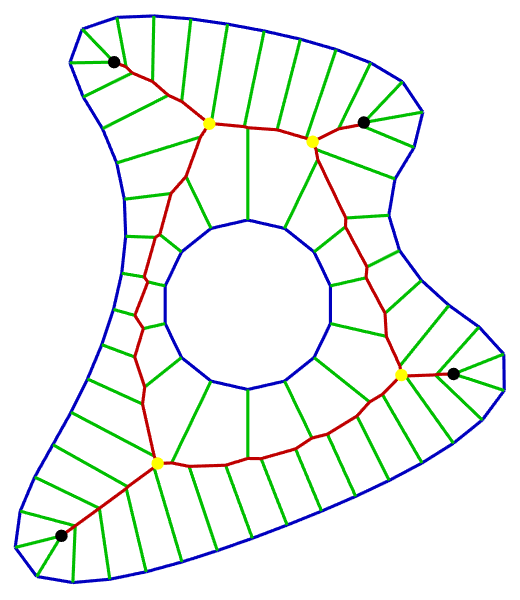}
\caption{Voronoi decomposition.   }
\label{fig:Results/StarwithHole/01}
    \end{subfigure}
        \hfill
    \begin{subfigure}[t]{0.24\linewidth}
\includegraphics[width=\linewidth]{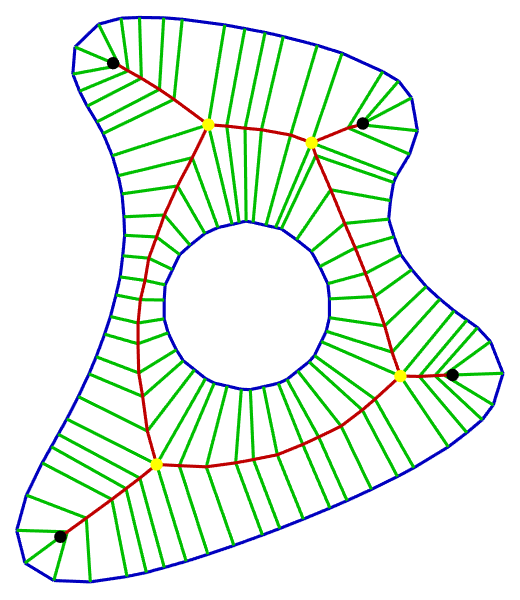}
\caption{Remeshing.     }
\label{fig:Results/StarwithHole/02}
    \end{subfigure}
        \hfill
    \begin{subfigure}[t]{0.24\linewidth}
\includegraphics[width=\linewidth]{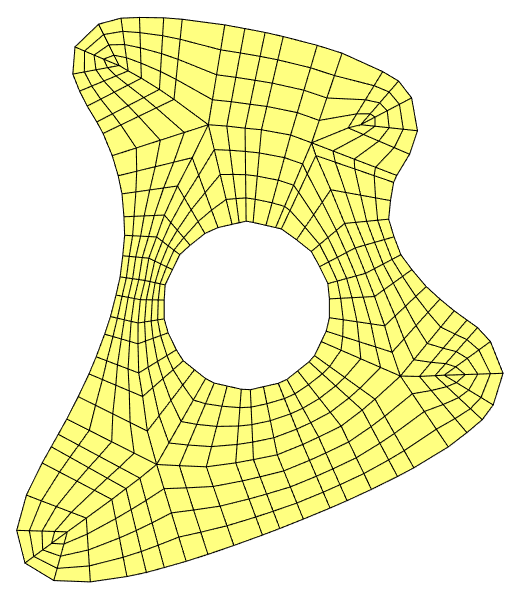}
\caption{Parametrization. }
\label{fig:Results/StarwithHole/03}
    \end{subfigure}
        \hfill
    \caption{The process of parametrization of the star-shaped domain with a circular hole. It is counterintuitive but the non-smooth medial axis is caused by a higher-resolution sample of the ellipse.  This creates local paraboloid regions that are usually caused by the Voronoi boundary of one line and one single point.  }
    \label{fig:StarwithHole}

  \centering
      \begin{subfigure}[t]{0.24\linewidth}
\includegraphics[width=\linewidth]{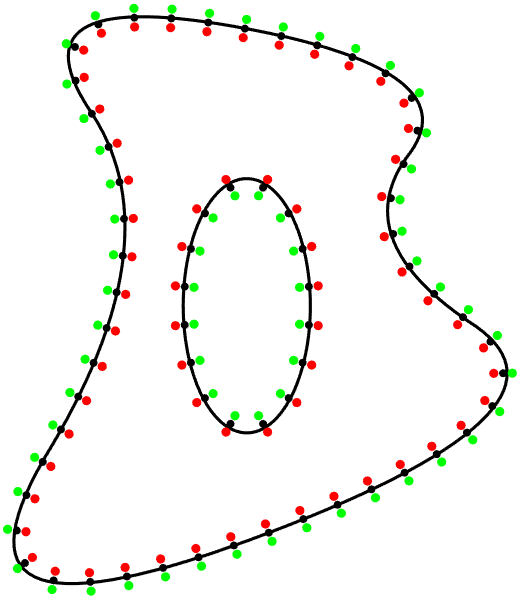}
\caption{Voronoi Sites: Dipoles.  }
\label{fig:Results/StarwithEllipticHole/00}
    \end{subfigure}
        \hfill
    \begin{subfigure}[t]{0.24\linewidth}
\includegraphics[width=\linewidth]{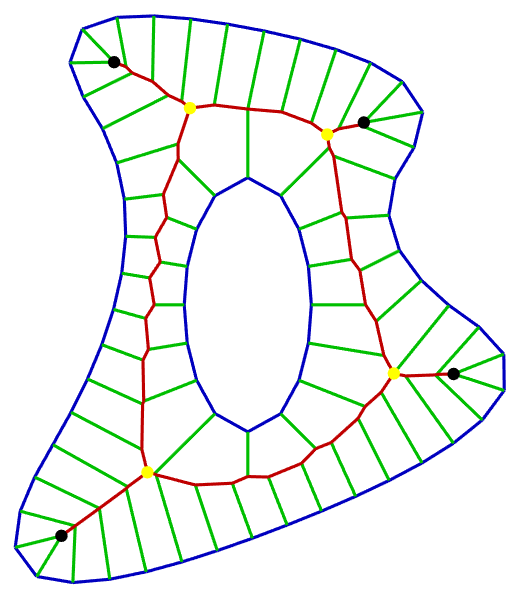}
\caption{Voronoi decomposition.  }
\label{fig:Results/StarwithEllipticHole/01}
    \end{subfigure}
        \hfill
    \begin{subfigure}[t]{0.24\linewidth}
\includegraphics[width=\linewidth]{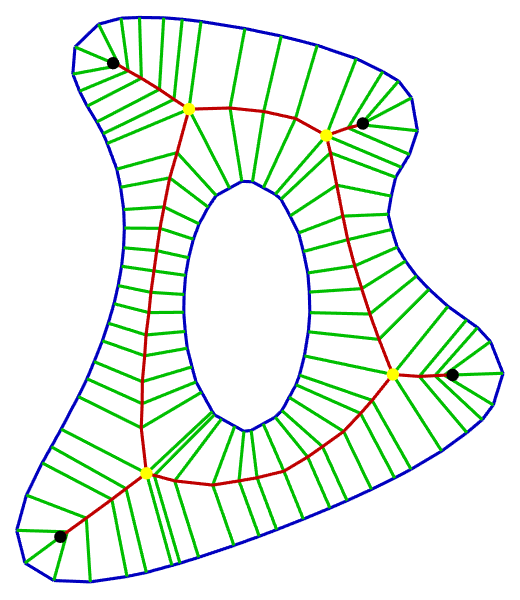}
\caption{Remeshing.  }
\label{fig:Results/StarwithEllipticHole/02}
    \end{subfigure}
        \hfill
    \begin{subfigure}[t]{0.24\linewidth}
\includegraphics[width=\linewidth]{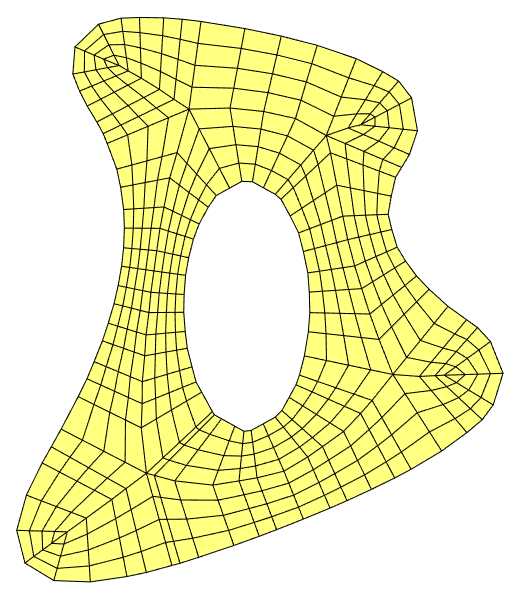}
\caption{Parametrization.  }
\label{fig:Results/StarwithEllipticHole/03}
    \end{subfigure}
        \hfill
    \caption{The process of parametrization of the star-shaped domain with an elliptic hole. Note that the medial axis is smoothed after the re-meshing.  }
    \label{fig:StarwithEllipticHole}
\end{figure*}

\begin{figure*}[htbp!]

  \centering
      \begin{subfigure}[t]{0.24\linewidth}
\includegraphics[width=\linewidth]{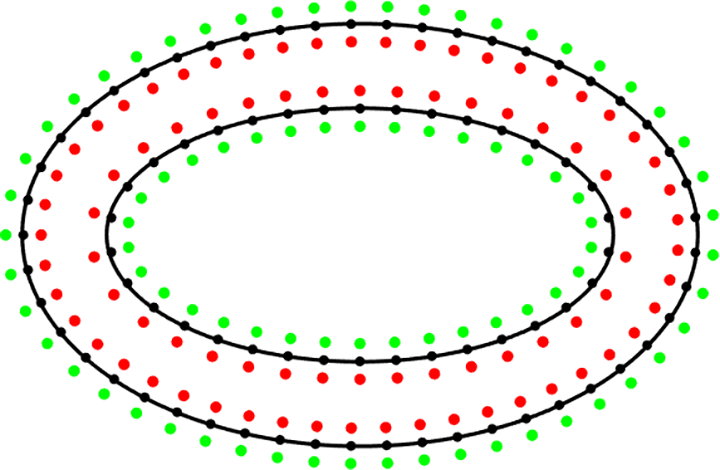}
\caption{Voronoi Sites: Dipoles. }
\label{fig:Results/Ellipse00degreeHole/00}
    \end{subfigure}
        \hfill
    \begin{subfigure}[t]{0.24\linewidth}
\includegraphics[width=\linewidth]{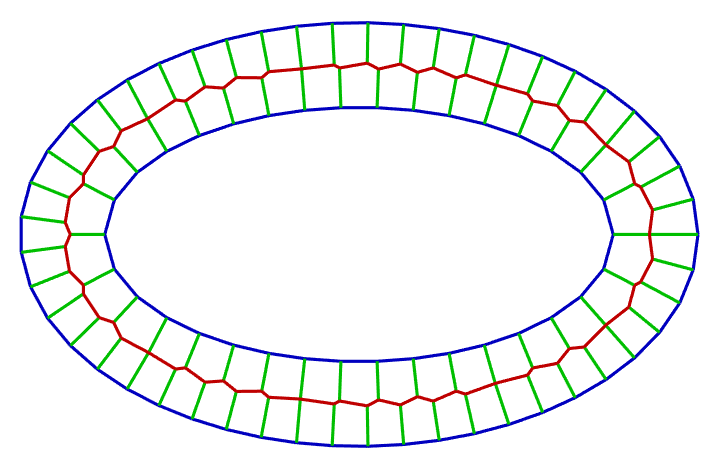}
\caption{Voronoi decomposition.   }
\label{fig:Results/Ellipse00degreeHole/01}
    \end{subfigure}
        \hfill
    \begin{subfigure}[t]{0.24\linewidth}
\includegraphics[width=\linewidth]{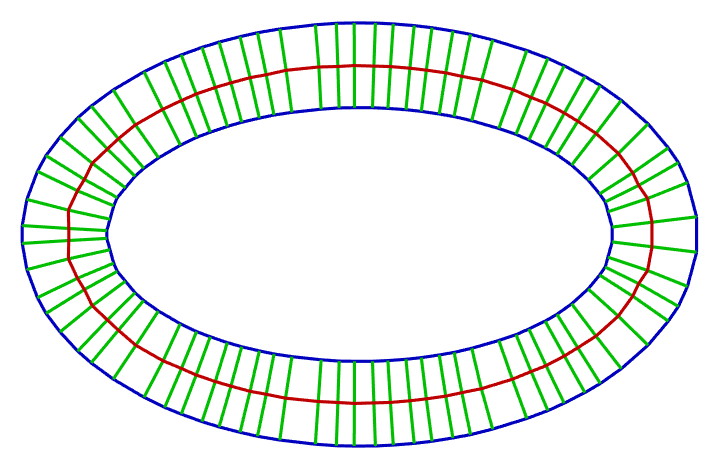}
\caption{Remeshing.     }
\label{fig:Results/Ellipse00degreeHole/02}
    \end{subfigure}
        \hfill
    \begin{subfigure}[t]{0.24\linewidth}
\includegraphics[width=\linewidth]{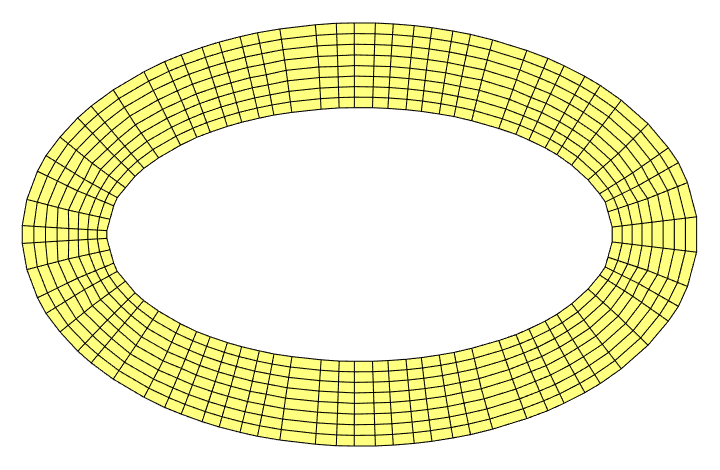}
\caption{Parametrization. }
\label{fig:Results/Ellipse00degreeHole/03}
    \end{subfigure}
        \hfill
    \caption{The process of parametrization of an Ellipsoidal domain with an ellipsoid-shaped hole. }
    \label{fig:Ellipse00degreeHole}
 
 \centering
      \begin{subfigure}[t]{0.24\linewidth}
\includegraphics[width=\linewidth]{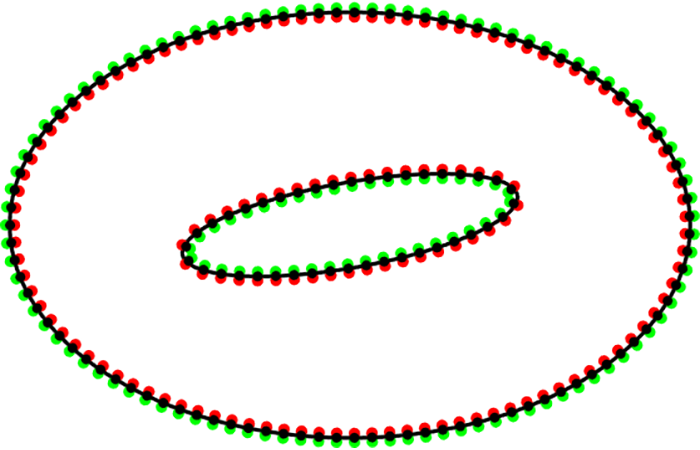}
\caption{Voronoi Sites: Dipoles. }
\label{fig:Results/Ellipse10degreeHole/00}
    \end{subfigure}
        \hfill
    \begin{subfigure}[t]{0.24\linewidth}
\includegraphics[width=\linewidth]{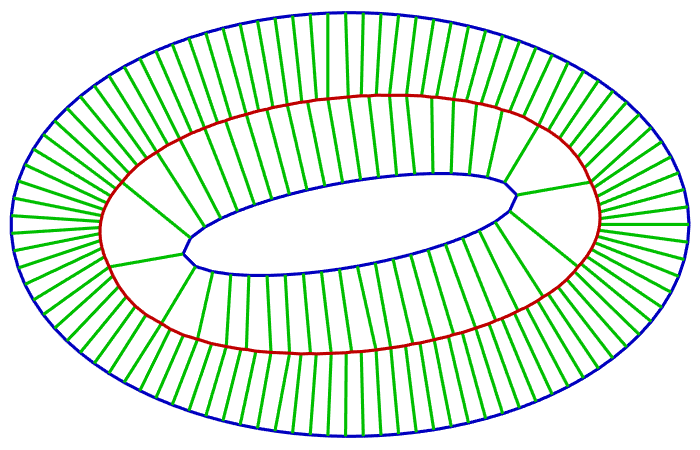}
\caption{Voronoi decomposition.   }
\label{fig:Results/Ellipse10degreeHole/01}
    \end{subfigure}
        \hfill
    \begin{subfigure}[t]{0.24\linewidth}
\includegraphics[width=\linewidth]{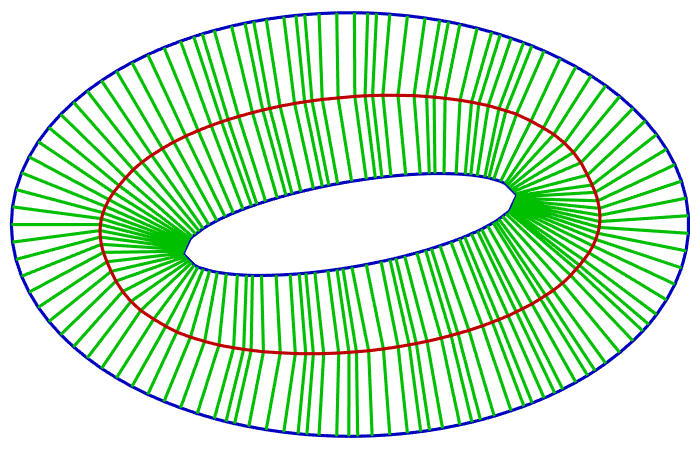}
\caption{Remeshing.     }
\label{fig:Results/Ellipse10degreeHole/02}
    \end{subfigure}
        \hfill
    \begin{subfigure}[t]{0.24\linewidth}
\includegraphics[width=\linewidth]{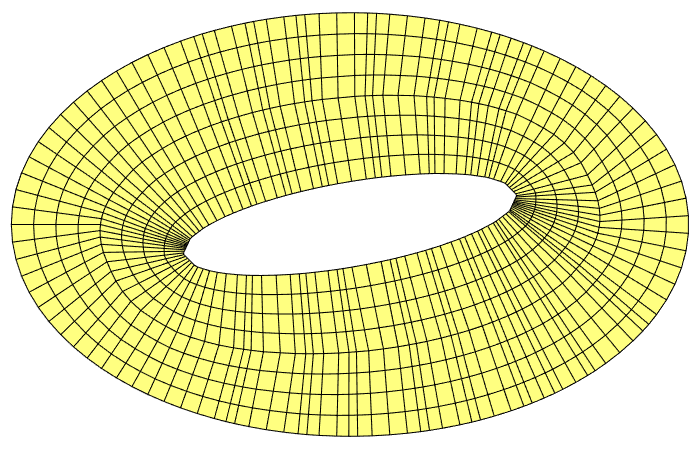}
\caption{Parametrization. }
\label{fig:Results/Ellipse10degreeHole/03}
    \end{subfigure}
        \hfill
    \caption{The process of parametrization of an  Ellipsoidal domain with 10 10-degree rotated ellipsoidal hole. }
    \label{fig:Ellipse10degreeHole}

  \centering
      \begin{subfigure}[t]{0.24\linewidth}
\includegraphics[width=\linewidth]{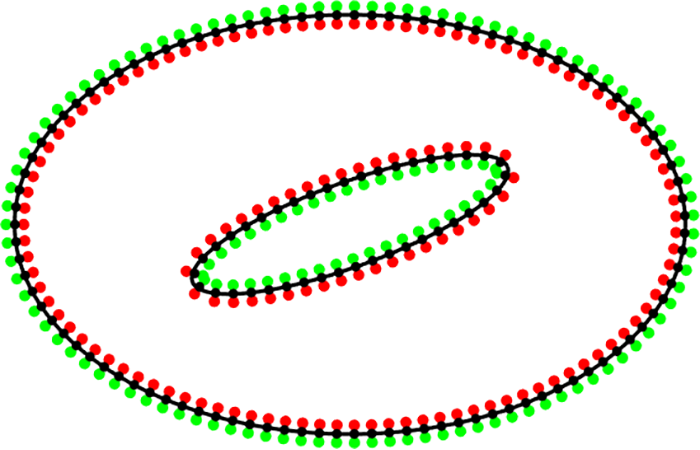}
\caption{Voronoi Sites: Dipoles. }
\label{fig:Results/Ellipse20degreeHole/00}
    \end{subfigure}
        \hfill
    \begin{subfigure}[t]{0.24\linewidth}
\includegraphics[width=\linewidth]{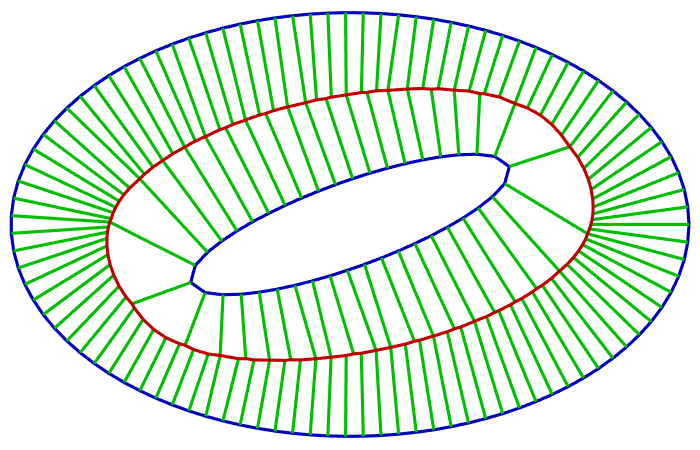}
\caption{Voronoi decomposition.   }
\label{fig:Results/Ellipse20degreeHole/01}
    \end{subfigure}
        \hfill
    \begin{subfigure}[t]{0.24\linewidth}
\includegraphics[width=\linewidth]{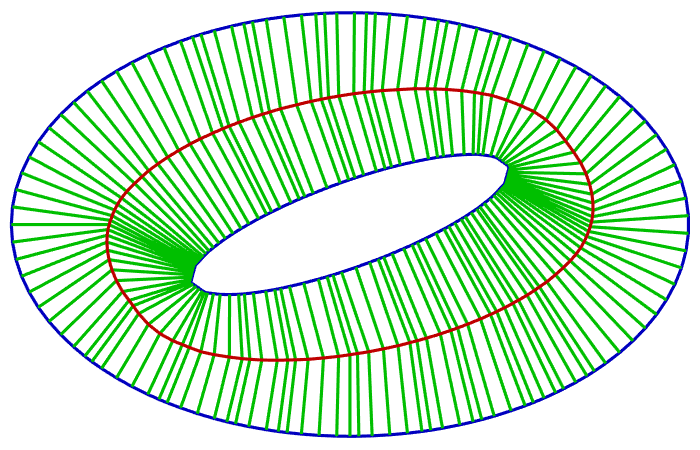}
\caption{Remeshing.     }
\label{fig:Results/Ellipse20degreeHole/02}
    \end{subfigure}
        \hfill
    \begin{subfigure}[t]{0.24\linewidth}
\includegraphics[width=\linewidth]{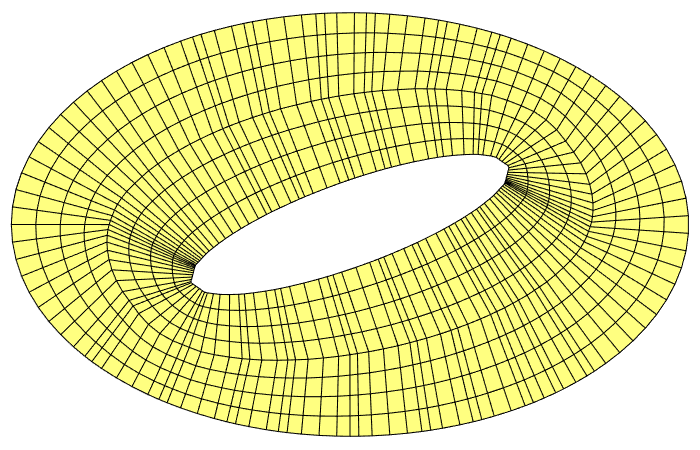}
\caption{Parametrization. }
\label{fig:Results/Ellipse20degreeHole/03}
    \end{subfigure}
        \hfill
    \caption{The process of parametrization of the ellipsoidal domain with a 20-degree rotated ellipsoidal hole. }
    \label{fig:Ellipse20degreeHole}

  \centering
      \begin{subfigure}[t]{0.24\linewidth}
\includegraphics[width=\linewidth]{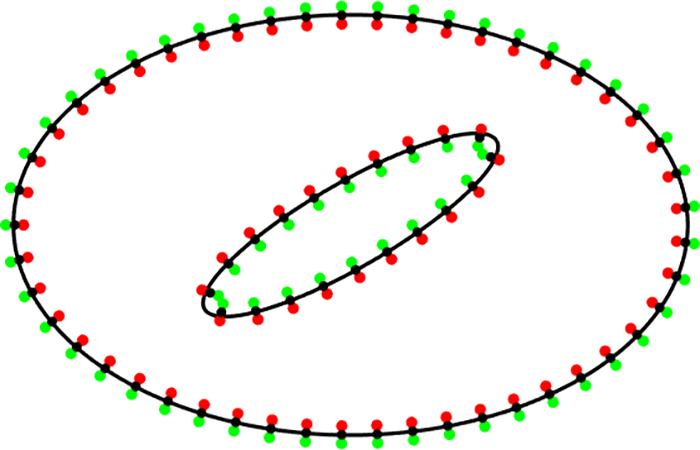}
\caption{Voronoi Sites: Dipoles.  }
\label{fig:Results/Ellipse30degreeHole/00}
    \end{subfigure}
        \hfill
    \begin{subfigure}[t]{0.24\linewidth}
\includegraphics[width=\linewidth]{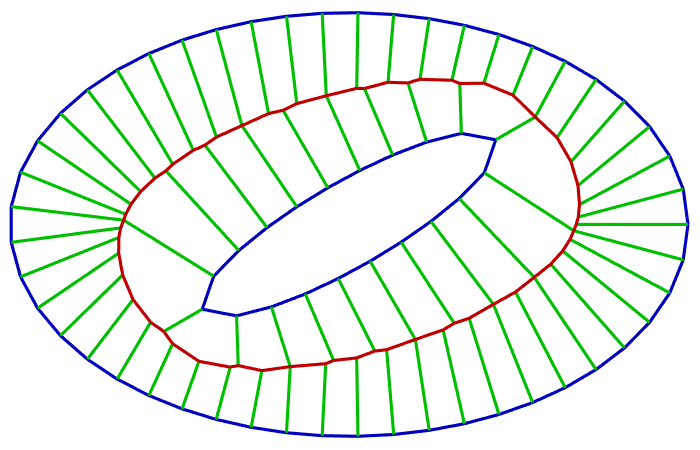}
\caption{Voronoi decomposition.  }
\label{fig:Results/Ellipse30degreeHole/01}
    \end{subfigure}
        \hfill
    \begin{subfigure}[t]{0.24\linewidth}
\includegraphics[width=\linewidth]{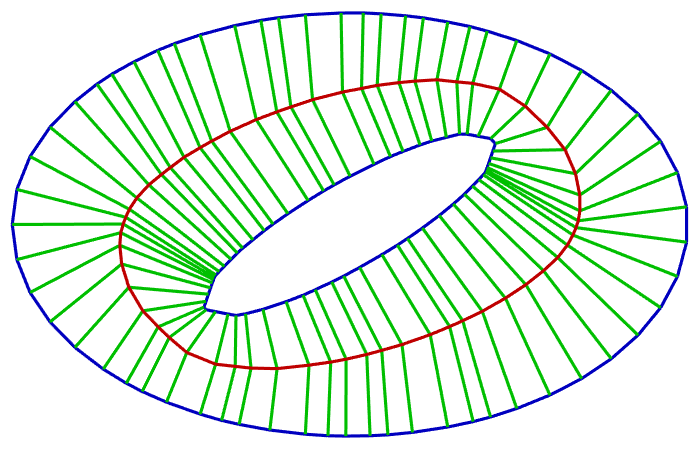}
\caption{Remeshing.  }
\label{fig:Results/Ellipse30degreeHole/02}
    \end{subfigure}
        \hfill
    \begin{subfigure}[t]{0.24\linewidth}
\includegraphics[width=\linewidth]{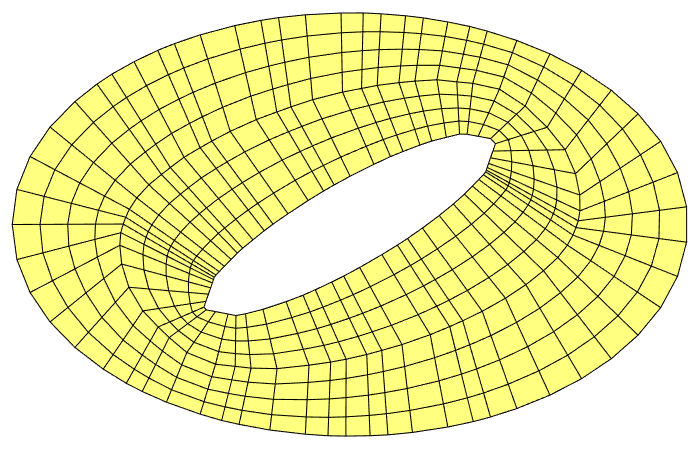}
\caption{Parametrization.  }
\label{fig:Results/Ellipse30degreeHole/03}
    \end{subfigure}
        \hfill
    \caption{The process of parametrization of the Ellipse shaped domain domain with a 30-degree rotated elliptic hole. }
    \label{fig:Ellipse30degreeHole}

  \centering
      \begin{subfigure}[t]{0.24\linewidth}
\includegraphics[width=\linewidth]{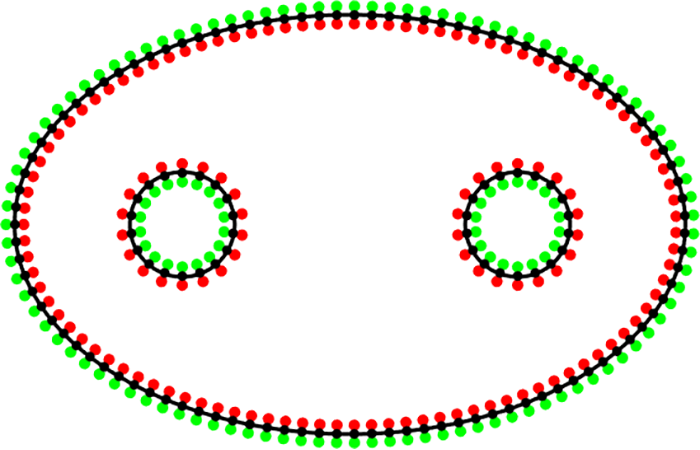}
\caption{Voronoi Sites: Dipoles.  }
\label{fig:Results/Ellipse2Holes/00}
    \end{subfigure}
        \hfill
    \begin{subfigure}[t]{0.24\linewidth}
\includegraphics[width=\linewidth]{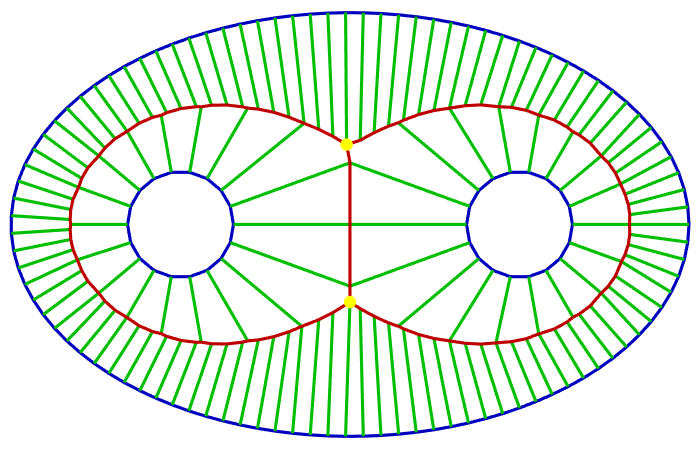}
\caption{Voronoi decomposition.  }
\label{fig:Results/Ellipse2Holes/01}
    \end{subfigure}
        \hfill
    \begin{subfigure}[t]{0.24\linewidth}
\includegraphics[width=\linewidth]{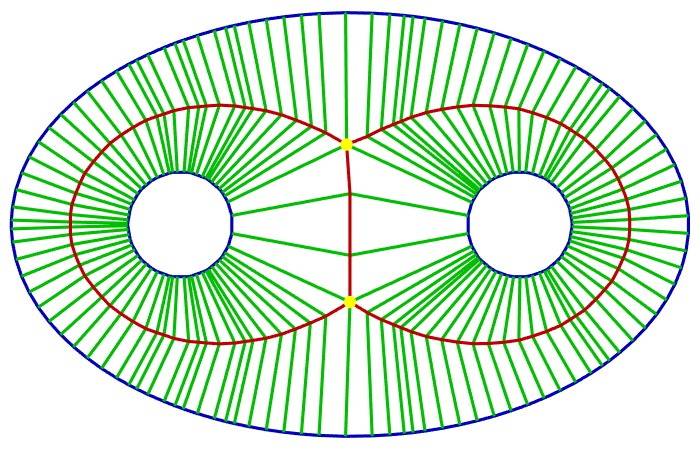}
\caption{Remeshing.  }
\label{fig:Results/Ellipse2Holes/02}
    \end{subfigure}
        \hfill
    \begin{subfigure}[t]{0.24\linewidth}
\includegraphics[width=\linewidth]{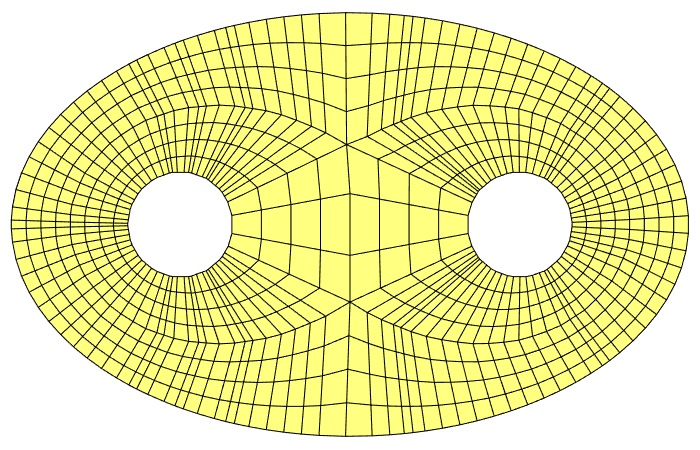}
\caption{Parametrization.  }
\label{fig:Results/Ellipse2Holes/03}
    \end{subfigure}
        \hfill
    \caption{The process of parametrization of the Ellipse shaped domain domain with two circular holes. }
    \label{fig:Ellipse2Holes}
    
\end{figure*}

In general, the corner-cutting induced by our interpolative re-meshing is clearly evident in all our examples wherein both the spine and the boundary are smoothed out. As expected, our parametrization results in quad-dominant meshing with triangular faces exclusively on the polar vertices of the spine (Figure~\ref{fig:Results/Star/03}). In general, we found that the method is also robust to holes regardless of their relative shapes (Figures \ref{fig:StarwithHole}, \ref{fig:StarwithEllipticHole}), sizes (Figure \ref{fig:Smiley}), or number (Figure \ref{fig:Ellipse2Holes}, \ref{fig:L3Holes}). Specially for our experiments with ellipses (Figures \ref{fig:Ellipse00degreeHole}, \ref{fig:Ellipse10degreeHole}, and \ref{fig:Ellipse30degreeHole}), we observe a visually continuous transition as the angle of the elliptical hole increases. Our method, as expected, is also agnostic to multiple components (Figure \ref{fig:SMI}).

Contrary to common intuition, we found that the algorithm was more robust for low sampling rates on the curves. One possible reason is that the number of near-zero edges increases as the curve tends to its limiting case because of the high sampling rate. This effect shows for all cases with holes wherein, the radial direction (limbs) is densely packed as we move from the boundary to the hole (Figure \ref{fig:RE_SAMPLING} left panel, \ref{fig:Smiley}). Another important artifact of sampling with an equal number of points is a lower quality in the medial axis (Figure \ref{fig:RE_SAMPLING} right panel). In such cases, using length-dependent sampling produces a better medial axis. The iso-parametric contours generated by our method (Figure \ref{fig:RE_CONTOUR}) demonstrate the manifestation of tangent discontinuities along the limbs connected to the branch vertices. While it is natural for this artifact to occur, this can be remedied with known regularization methods. What is important here is that the explicit knowledge of the location of these discontinuities allows for an easy way to localize any regularization thereby preserving the parametrization in the remaining domain.

These experiments give us three-step results (1) Voronoi decomposition (radial and medial axis),(2) Remeshing, and (3) Parametrization. The remeshing process turns the 3-valence points into 4-valence points and non-quadrilateral cells into a combination of triangular and quadrilateral shapes.
Parametrization mostly gives us quadrilateral cells. Additionally,we observe triangular cells as shown in Figure~\ref{fig:Results/Star/03} as we move from the boundary towards the medial axis near the corner region of these domains. This can be observed for the domains bounded by a circular arc. We also observe that rhombus-shaped cells of the medial axis have multiple branches (see figure~\ref{fig:Results/Star/03}).
Contours are represented by a wide range of closed curves which have a single continuous medial axis and medial axis with branches. We observe a dense distribution of the curves near sharp curvatures.

We sample points equidistantly on the input curve. As we increase the number of sample points, the possibility of generating very small edges increases as a result of Voronoi decomposition because of numerical inaccuracies. These edges pose a challenge in remeshing.
Our algorithm carefully identifies and removes these edges without losing the information about the cell structure. 
This equidistant sampling poses a challenge for domains generated from multiple closed curves (see Figure~\ref{fig:Results/Smiley/00}). Using the same number of sampling points to sample equidistant points on all the curves may bring in some numerical inaccuracies because the arc lengths of each curve are different. This can be resolved if we sample individual curves based on the lengths of individual curves. Remeshing of the initial Voronoi decomposition also allows us to smooth the zig-zag pattern of the medial axis generated originally from Voronoi.This is implicitly done by corner cutting of the medial axis.

\section{Discussion}

\subsection{Medial Parametrization \& Implicts}
\begin{figure*}
  \centering
      \begin{subfigure}[t]{0.24\linewidth}
\includegraphics[width=\linewidth]{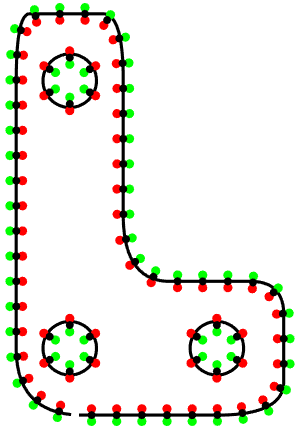}
\caption{Voronoi Sites: Dipoles.  }
\label{fig:Results/L3Holes/00}
    \end{subfigure}
        \hfill
    \begin{subfigure}[t]{0.24\linewidth}
\includegraphics[width=\linewidth]{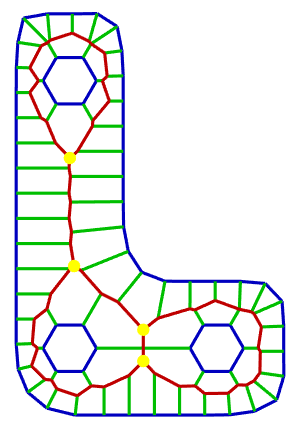}
\caption{Voronoi decomposition.  }
\label{fig:Results/L3Holes/01}
    \end{subfigure}
        \hfill
    \begin{subfigure}[t]{0.24\linewidth}
\includegraphics[width=\linewidth]{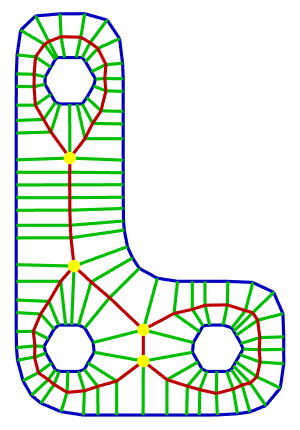}
\caption{Remeshing.  }
\label{fig:Results/L3Holes/02}
    \end{subfigure}
        \hfill
    \begin{subfigure}[t]{0.24\linewidth}
\includegraphics[width=\linewidth]{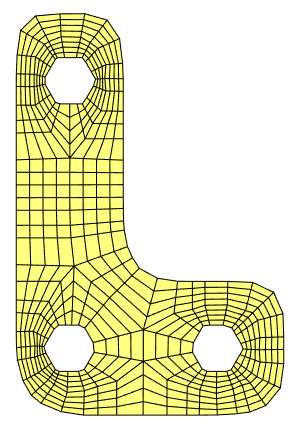}
\caption{Parametrization.  }
\label{fig:Results/L3Holes/03}
    \end{subfigure}
        \hfill
    \caption{The process of parameterization of the L-shaped domain with three circular holes. }
    \label{fig:L3Holes}
    \end{figure*}

The connection between Voronoi partition, medial axis, and implicit curves and surfaces \cite{turk2002modelling}, though deep and interesting, is not entirely surprising. Our formulation defines an implicit function $F: \mathbb{R}^2 \rightarrow \mathbb{R}$ in the following form: $$F(\mathbf{P}) = min( | \mathbf{P} - \mathbf{P}^{(+)}_{i}|) - min( | \mathbf{P} - \mathbf{P}^{(-)}_{i}|)$$
where $min()$ is the minimum value of each Euclidean distance function $|.|)$ for all $i$ in $i=0,1,\ldots, N-1$. The implicit curves are defined using this function as follows
$$C = \{ \mathbf{P} \; \;  |\; \;  F(\mathbf{P}) \; \;  =\; \;  0  \} $$
and the whole domain is given as 
$$D = \{ \mathbf{P} \; \;  |\; \;  F(\mathbf{P}) \; \; \leq \; \; 0  \} $$
The function $F(\mathbf{P})$ in the domain is actually the parametrization. When the number of dipoles goes to infinity this function approaches the exact distance from the original curve. 
Even a relatively low number of samples is sufficient to obtain a reasonable approximation of this function. The re-meshing process simultaneously smooths the medial axis and boundary curves by removing zigzag structures caused by Voronoi tessellation (Figure~\ref{fig:L3Holes}). 

\subsection{Meshing}
An important property of this parametrization is that it consists of 1D linear components. Therefore, it can be straightforward to define more complicated functions using this parametrization. Using this extension, it can be possible to develop alternative methods for higher-order meshing of curved 2D domains \cite{mandad2020bezier}. These functions can be specifically used in GIS applications to produce landscapes with desired properties. One constraint, on the other hand, is that most branch vertices in the medial axis are extraordinary type, i.e. not 4-valence vertices. It is impossible to avoid $G^2$ discontinuities. Moreover, polar vertices have 1-valence, which also requires care.

\subsection{Alternate Computation Strategies}
Our algorithm mainly revolved around the generation of 4-valency vertices leading to quad-dominant domain partition. Our specific choice of the interpolative strategy was by and large topological and relied less on the geometric and therefore numerical challenges that are inherent in Voronoi tessellations. This allowed for a robust implementation of medial parametrization. Secondly, the interpolation naturally lent itself to automatic and simultaneous smoothing of the medial axis and the boundary without the need for any post-processing. Having said that, there may be other possible ways to generate quad-dominant partitions. For instance, an alternative would be to simply add new limbs at each of the existing spine vertices and subsequently smooth the spine using existing regularization methods, the simplest being Laplacian smoothing. Another potential approach would be to simplify the medial axis as a set of as straight as possible segments and produce limbs through the closest point computation from the boundary to the spine. 
\section{Conclusion and Future Directions}

In this paper, we introduced a new parametrization, which we call medial parametrization. Medial parametrization allows for the description of points contained in any arbitrarily shaped compact planar domain bounded by a set of simple closed curves. Especially for domains homeomorphic to a disk, medial parametrization turns into a generalization of polar coordinates. We further developed a method to construct an approximation of this parametrization by using Voronoi tessellation.

Medial parametrization provides a wide range of potential future works. We will first list 2D applications. This parameterization can be used to create generalized Archimedean spirals, which are useful for creating paths for 3D printing. Archimedean spirals can also be used to construct a remeshing that appears to be a mosaic. Note that in mosaics, shapes have to be square shaped actual topology can be a hexagon or pentagon with a large number of 3-valent vertices similar to Igloo blocks. 

We see several applications of our approach ranging from topographical modeling in geographical information systems (GIS), the design of architectural geometry, and texture mapping. For instance, the method can also be extended to include open curves, trees, and even graphs (a network of curves) in fundamental domains of wallpaper symmetries. Since fundamental domains are compact, it can be straightforward to extend this method to wallpaper cases. On the other hand, to handle a network of curves, there is a need for extension of the dipole concept to obtain multi-colored regions. Using open curves and fundamental domains can especially be useful for GIS applications, in which we may want to control the landscape features which can include tree-line features such as rivers. This extension can also \textcolor{red}{lends} itself parametrization of surfaces. The network of curves can also be used as an interface to design architectural free-form surfaces to construct curved developable surfaces \cite{eigensatz2010paneling}. 

The method can also be extended to include open curves, trees, and even graphs (a network of curves) in fundamental domains of wallpaper symmetries. Since fundamental domains are compact, it can be straightforward to extend this method to wallpaper cases. On the other hand, to handle a network of curves, there is a need for an extension of the dipole concept to obtain multi-colored regions. Using open curves and fundamental domains can especially be useful for GIS applications, in which we may want to control the landscape features which can include tree-line features such as rivers. This extension can also lend itself parametrization of surfaces. The network of curves can also be used as an interface to design architectural free-form surfaces to construct curved developable surfaces \cite{eigensatz2010paneling}. 

In principle, this approach can also be applied to 2-manifold surfaces with a set of closed curves on the surface. Computing Voronoi partition on the surface \cite{liu2010construction} could provide new ways to obtain surface parametrization for quad-dominant re-meshing. Using this, we also envision possibilities for parameterizing compact volumetric domains that are bounded by surfaces. Once we obtain surface parametrization using Voronoi tessellations with geodesic distance, it may be possible to use the resulting structure to produce dipoles in 3D to construct a parametrization in 3D space. Overall, we believe envision medial parametrization as a fundamental tool in the rich tool-set for geometric computing and analysis and will lead to rich research directions in the future.


\bibliographystyle{unsrtnat}
\bibliography{references}

\begin{thebibliography}{31}
\providecommand{\natexlab}[1]{#1}
\providecommand{\url}[1]{\texttt{#1}}
\expandafter\ifx\csname urlstyle\endcsname\relax
  \providecommand{\doi}[1]{doi: #1}\else
  \providecommand{\doi}{doi: \begingroup \urlstyle{rm}\Url}\fi

\bibitem[Sheffer et~al.(2006)Sheffer, Praun, and Rose]{Sheffer2006}
Alla Sheffer, Emil Praun, and Kenneth Rose.
\newblock Mesh parameterization methods and their applications.
\newblock \emph{Found. Trends. Comput. Graph. Vis.}, 2\penalty0 (2):\penalty0 105–171, January 2006.
\newblock ISSN 1572-2740.
\newblock \doi{10.1561/0600000011}.
\newblock URL \url{https://doi.org/10.1561/0600000011}.

\bibitem[Farin(2002)]{FARIN20021}
Gerald Farin.
\newblock Chapter 1 - a history of curves and surfaces in cagd.
\newblock In Gerald Farin, Josef Hoschek, and Myung-Soo Kim, editors, \emph{Handbook of Computer Aided Geometric Design}, pages 1--21. North-Holland, Amsterdam, 2002.
\newblock ISBN 978-0-444-51104-1.
\newblock \doi{https://doi.org/10.1016/B978-044451104-1/50002-2}.
\newblock URL \url{https://www.sciencedirect.com/science/article/pii/B9780444511041500022}.

\bibitem[Coxeter(1961)]{coxeter1961introduction}
Harold Scott~Macdonald Coxeter.
\newblock \emph{Introduction to geometry}.
\newblock New York, London, 1961.

\bibitem[Bartels et~al.(1995)Bartels, Beatty, and Barsky]{bartels1995introduction}
Richard~H Bartels, John~C Beatty, and Brian~A Barsky.
\newblock \emph{An introduction to splines for use in computer graphics and geometric modeling}.
\newblock Morgan Kaufmann, 1995.

\bibitem[Wachspress(1975)]{wachspress1975rational}
Eugene~L. Wachspress.
\newblock A rational finite element basis.
\newblock \emph{Academic Press}, 114, 1975.

\bibitem[Gregory et~al.(1990)Gregory, Lau, and Zhou]{gregory1990smooth}
John~A Gregory, Vincent~KH Lau, and Jianwei Zhou.
\newblock Smooth parametric surfaces and n-sided patches.
\newblock In \emph{Computation of curves and surfaces}, pages 457--498. Springer, 1990.

\bibitem[Loop and DeRose(1990)]{loop1990generalized}
Charles Loop and Tony~D DeRose.
\newblock Generalized b-spline surfaces of arbitrary topology.
\newblock In \emph{Proceedings of the 17th annual conference on Computer graphics and interactive techniques}, pages 347--356, 1990.

\bibitem[Loop and DeRose(1989)]{loop1989multisided}
Charles~T Loop and Tony~D DeRose.
\newblock A multisided generalization of b{\'e}zier surfaces.
\newblock \emph{ACM Transactions on Graphics (TOG)}, 8\penalty0 (3):\penalty0 204--234, 1989.

\bibitem[Floater(2003)]{floater2003mean}
Michael~S Floater.
\newblock Mean value coordinates.
\newblock \emph{Computer aided geometric design}, 20\penalty0 (1):\penalty0 19--27, 2003.

\bibitem[Rustamov et~al.(2009)Rustamov, Lipman, and Funkhouser]{rustamov2009interior}
Raif~M Rustamov, Yaron Lipman, and Thomas Funkhouser.
\newblock Interior distance using barycentric coordinates.
\newblock \emph{Computer Graphics Forum}, 28\penalty0 (5):\penalty0 1279--1288, 2009.

\bibitem[Ju et~al.(2023)Ju, Schaefer, and Warren]{ju2005mean}
Tao Ju, Scott Schaefer, and Joe Warren.
\newblock Mean value coordinates for closed triangular meshes.
\newblock In \emph{Seminal Graphics Papers: Pushing the Boundaries, Volume 2: Originally published in ACM Siggraph 2005}, pages 223--228. ACM Siggraph, 2023.

\bibitem[Robinson et~al.(1992{\natexlab{a}})Robinson, Griffin, and Colchester]{robinson1992delaunay}
Glynn Robinson, Lewis Griffin, and Alan Colchester.
\newblock The delaunay/voronoi selection graph: a method for extracting shape information from 2-d dot-patterns with an extension to 3-d.
\newblock In \emph{BMVC92}, pages 19--28. Springer, 1992{\natexlab{a}}.

\bibitem[Ogniewicz(1994)]{ogniewicz1994skeleton}
RL~Ogniewicz.
\newblock Skeleton-space: a multiscale shape description combining region and boundary information.
\newblock \emph{age}, 7\penalty0 (8):\penalty0 9, 1994.

\bibitem[Dey and Zhao(2004)]{dey2004approximating}
Tamal~K Dey and Wulue Zhao.
\newblock Approximating the medial axis from the voronoi diagram with a convergence guarantee.
\newblock \emph{Algorithmica}, 38\penalty0 (1):\penalty0 179--200, 2004.

\bibitem[Sharma et~al.(2006)Sharma, Mioc, and Anton]{sharma2006voronoi}
Ojaswa Sharma, Darka Mioc, and Francois Anton.
\newblock Voronoi diagram based automated skeleton extraction from colour scanned maps.
\newblock In \emph{2006 3rd International Symposium on Voronoi Diagrams in Science and Engineering}, pages 186--195. IEEE, 2006.

\bibitem[Beristain et~al.(2012)Beristain, Gra{\~n}a, and Gonzalez]{beristain2012pruning}
Andoni Beristain, Manuel Gra{\~n}a, and Ana~I Gonzalez.
\newblock A pruning algorithm for stable voronoi skeletons.
\newblock \emph{Journal of Mathematical Imaging and Vision}, 42\penalty0 (2):\penalty0 225--237, 2012.

\bibitem[Liu et~al.(2012)Liu, Wu, Hsu, Peterson, and Xu]{liu2012generation}
Hongzhi Liu, Zhonghai Wu, D~Frank Hsu, Bradley~S Peterson, and Dongrong Xu.
\newblock On the generation and pruning of skeletons using generalized voronoi diagrams.
\newblock \emph{Pattern Recognition Letters}, 33\penalty0 (16):\penalty0 2113--2119, 2012.

\bibitem[Amenta et~al.(1998{\natexlab{a}})Amenta, Bern, and Eppstein]{amenta1998crust}
Nina Amenta, Marshall Bern, and David Eppstein.
\newblock The crust and the $\beta$-skeleton: Combinatorial curve reconstruction.
\newblock \emph{Graphical models and image processing}, 60\penalty0 (2):\penalty0 125--135, 1998{\natexlab{a}}.

\bibitem[Amenta et~al.(1998{\natexlab{b}})Amenta, Bern, and Kamvysselis]{amenta1998new}
Nina Amenta, Marshall Bern, and Manolis Kamvysselis.
\newblock A new voronoi-based surface reconstruction algorithm.
\newblock In \emph{Proceedings of the 25th annual conference on Computer graphics and interactive techniques}, pages 415--421, 1998{\natexlab{b}}.

\bibitem[Robinson et~al.(1992{\natexlab{b}})Robinson, Colchester, Griffin, and Hawkes]{robinson1992integrated}
Glynn~P Robinson, Alan~CF Colchester, Lewis~D Griffin, and David~J Hawkes.
\newblock Integrated skeleton and boundary shape representation for medical image interpretation.
\newblock In \emph{European Conference on Computer Vision}, pages 725--729. Springer, 1992{\natexlab{b}}.

\bibitem[Gold(1999)]{gold1999crust}
Christopher Gold.
\newblock Crust and anti-crust: a one-step boundary and skeleton extraction algorithm.
\newblock In \emph{Proceedings of the fifteenth annual symposium on Computational geometry}, pages 189--196, 1999.

\bibitem[Gold and Snoeyink(2001)]{gold2001one}
Christopher Gold and Jack Snoeyink.
\newblock A one-step crust and skeleton extraction algorithm.
\newblock \emph{Algorithmica}, 30\penalty0 (2):\penalty0 144--163, 2001.

\bibitem[Itoh et~al.(1998)Itoh, Shimada, Inoue, Yamada, and Furuhata]{itoh1998automated}
Takayuki Itoh, Kenji Shimada, Keisuke Inoue, Atsushi Yamada, and Tomotake Furuhata.
\newblock Automated conversion of 2d triangular mesh into quadrilateral mesh with directionality control.
\newblock In \emph{IMR}, pages 77--86, 1998.

\bibitem[Marinov and Kobbelt(2006)]{marinov2006robust}
Martin Marinov and Leif Kobbelt.
\newblock A robust two-step procedure for quad-dominant remeshing.
\newblock \emph{Computer Graphics Forum}, 25\penalty0 (3):\penalty0 537--546, 2006.

\bibitem[Alliez et~al.(2008)Alliez, Ucelli, Gotsman, and Attene]{alliez2008recent}
Pierre Alliez, Giuliana Ucelli, Craig Gotsman, and Marco Attene.
\newblock Recent advances in remeshing of surfaces.
\newblock \emph{Shape analysis and structuring}, pages 53--82, 2008.

\bibitem[Tam and Armstrong(1991)]{tam19912d}
TKH Tam and Cecil~G Armstrong.
\newblock 2d finite element mesh generation by medial axis subdivision.
\newblock \emph{Advances in engineering software and workstations}, 13\penalty0 (5-6):\penalty0 313--324, 1991.

\bibitem[Sharma et~al.(2009)Sharma, Anton, and Mioc]{Sharma2009}
Ojaswa Sharma, François Anton, and Darka Mioc.
\newblock On the isomorphism between the medial axis and a dual of the delaunay graph.
\newblock In \emph{2009 Sixth International Symposium on Voronoi Diagrams}, pages 89--95, 2009.
\newblock \doi{10.1109/ISVD.2009.18}.

\bibitem[Turk and O'brien(2002)]{turk2002modelling}
Greg Turk and James~F O'brien.
\newblock Modelling with implicit surfaces that interpolate.
\newblock \emph{ACM Transactions on Graphics (TOG)}, 21\penalty0 (4):\penalty0 855--873, 2002.

\bibitem[Mandad and Campen(2020)]{mandad2020bezier}
Manish Mandad and Marcel Campen.
\newblock B{\'e}zier guarding: precise higher-order meshing of curved 2d domains.
\newblock \emph{ACM Transactions on Graphics (TOG)}, 39\penalty0 (4):\penalty0 103--1, 2020.

\bibitem[Eigensatz et~al.(2010)Eigensatz, Kilian, Schiftner, Mitra, Pottmann, and Pauly]{eigensatz2010paneling}
Michael Eigensatz, Martin Kilian, Alexander Schiftner, Niloy~J Mitra, Helmut Pottmann, and Mark Pauly.
\newblock Paneling architectural freeform surfaces.
\newblock In \emph{ACM SIGGRAPH 2010 papers}, pages 1--10. ACM Siggraph, New York, NY, USA, 2010.

\bibitem[Liu et~al.(2010)Liu, Chen, and Tang]{liu2010construction}
Yong-Jin Liu, Zhanqing Chen, and Kai Tang.
\newblock Construction of iso-contours, bisectors, and voronoi diagrams on triangulated surfaces.
\newblock \emph{IEEE Transactions on Pattern Analysis and Machine Intelligence}, 33\penalty0 (8):\penalty0 1502--1517, 2010.

\end{thebibliography}

\end{document}